\documentclass[aps,pra,twocolumn,groupedaddress,floatfix,showpacs,longbibliography]{revtex4-1}
\usepackage{graphicx}
\usepackage{amsmath}
\usepackage{color}

\begin{document}

\title{Hydrodynamic Relaxation in a Strongly Interacting Fermi Gas}

\author{Xin Wang, Xiang Li, Ilya Arakelyan, and J. E. Thomas}

\affiliation{$^{1}$Department of  Physics, North Carolina State University, Raleigh, NC 27695, USA}

\date{\today}

\begin{abstract}
We measure the free decay of a spatially periodic density profile in a normal fluid strongly interacting Fermi gas, which is confined in a box potential. This spatial profile is initially created in thermal equilibrium by a perturbing potential. After the perturbation is abruptly extinguished, the dominant spatial Fourier component exhibits an exponentially decaying (thermally diffusive) mode and a decaying oscillatory (first sound) mode, enabling independent measurement of the thermal conductivity and the shear viscosity directly from the time-dependent evolution.
\end{abstract}

\maketitle

Studies of thermodynamics and hydrodynamic transport in strongly correlated Fermi gases connect widely different forms of matter across vast energy scales~\cite{NJPReview,UrbanReview,BlochReview}. Strongly interacting Fermi gases are created by tuning a trapped, two-component cloud near a collisional (Feshbach) resonance~\cite{OHaraScience}.
A resonantly interacting or unitary Fermi gas is of special interest, as it is a scale-invariant, strongly interacting quantum many-body system, with thermodynamic and transport properties that are universal functions of the density and temperature~\cite{HoUniversalThermo}, permitting parameter-free comparisons with predictions.

However, the behavior of the shear viscosity $\eta$ and the thermal conductivity $\kappa_T$ of a unitary Fermi gas is not yet established. Measurement of hydrodynamic flow in freely expanding clouds~\cite{CaoViscosity,JosephShearNearSF} enables extraction of $\eta$ in the normal fluid regime, but requires a second order hydrodynamics model to properly account for ballistic flow in the dilute edges~\cite{BluhmSchaeferModIndep,BluhmSchaeferLocalViscosity}.  Recent measurements of the sound diffusivity~\cite{MZSound}, by sound attenuation in a driven, uniform density, unitary Fermi gas, constrain $\eta$ and $\kappa_T$, but they are not independently determined~\cite{MZTempWave}.

\begin{figure}[htb]
\hspace{-0.25in}\includegraphics[width=3.5in]{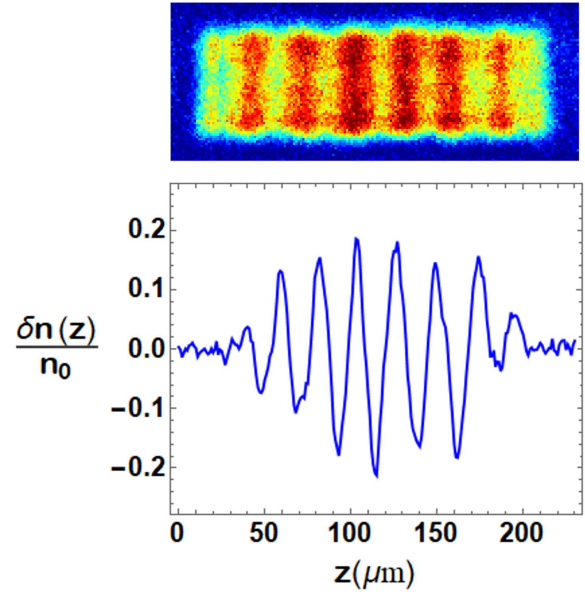}
\caption{A unitary Fermi gas is loaded into a box potential with a small static spatially periodic perturbation $\delta U$, creating a spatially periodic 1D density profile. After $\delta U$ is abruptly extinguished, the dominant Fourier component exhibits a two-mode  oscillatory decay (see Fig.~\ref{fig:Fourier}).
\label{fig:boxdensity}}
\end{figure}

\begin{figure}[h!]
\includegraphics[width=3.25in]{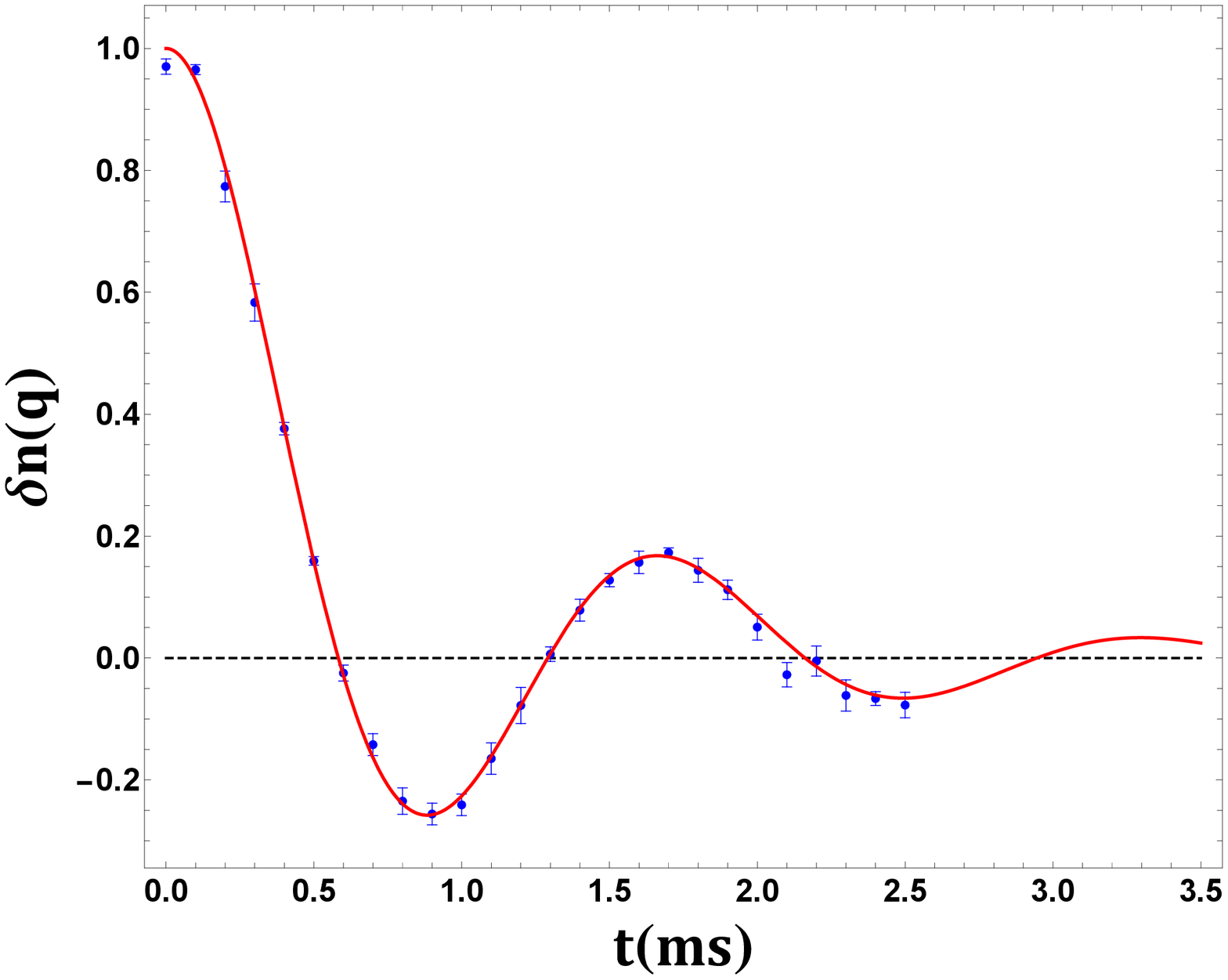}\\
\vspace{-2.55in}\hspace{1.4in}\includegraphics[width=1.75in]{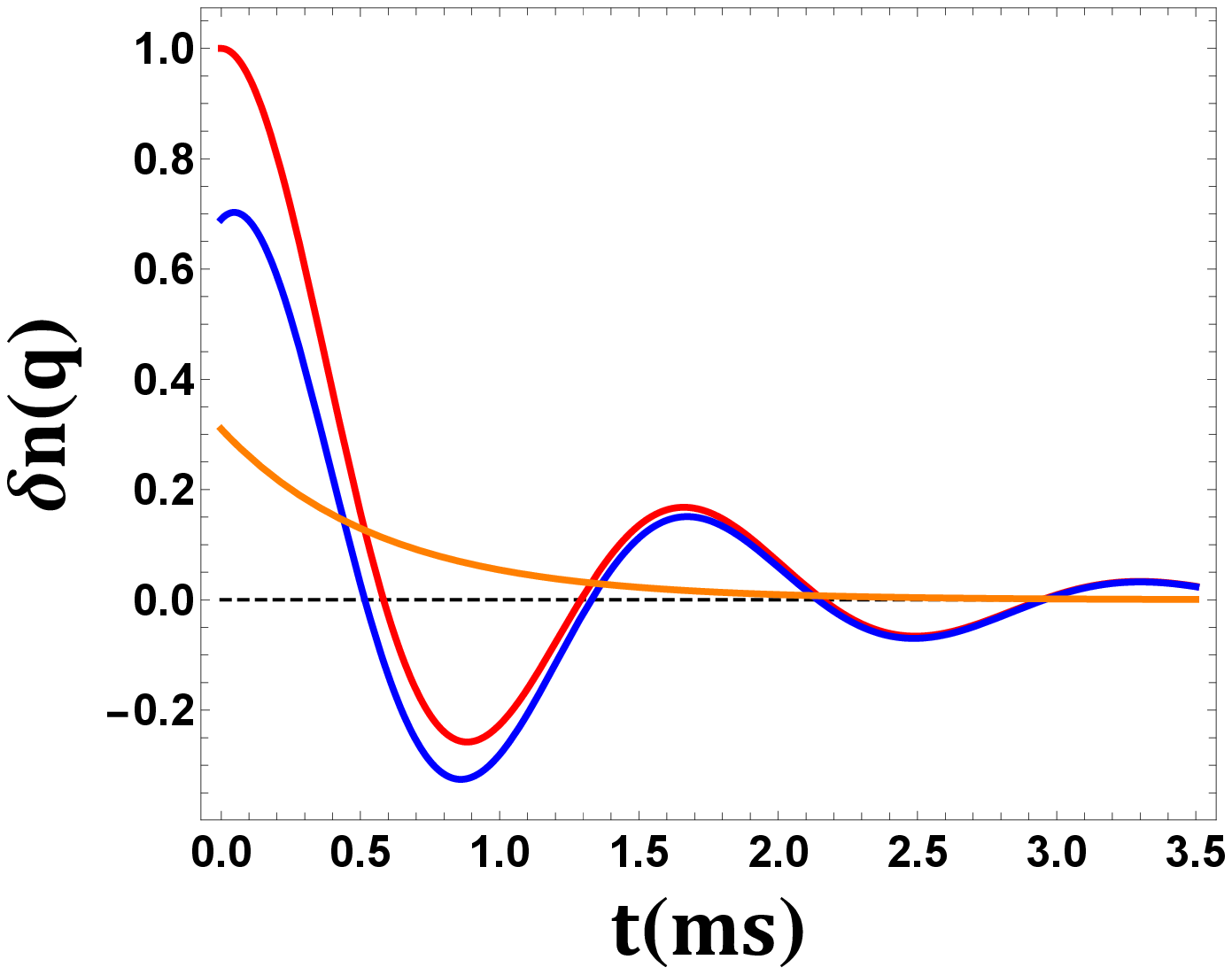}\\
\vspace{1.25in}\caption{Real part of the Fourier transform of the density perturbation $\delta n(q,t)$ for $q=2 \pi/\lambda$ with $\lambda = 22.7\,\mu$m. The reduced temperature $T/T_F =0.46$. Blue dots (data); Red curve: Analytic hydrodynamics model, eq.~\ref{eq:FTvsTime}. Inset shows contributions of thermal diffusion (orange exponential) and first sound (blue).  The error bars are the standard  deviation of the mean of $\delta n(q,t)$ for 5-8 runs,  taken in random time order.
\label{fig:Fourier}}
\end{figure}

In this Letter, we report new time-domain, free evolution methods for measuring hydrodynamic transport coefficients in a normal fluid unitary Fermi gas.  We confine a cloud of $^6$Li atoms in a repulsive box potential, producing a sample of nearly uniform density. A density perturbation is then created, Fig.~\ref{fig:boxdensity}, by applying a small static optical potential that is spatially periodic along one axis. After equilibrium is established, the perturbing potential is abruptly extinguished. We measure the time-dependence of the dominant spatial Fourier component of the density,  $\delta n(q,t)$, Fig.~\ref{fig:Fourier}, which exhibits an exponentially decaying mode that measures the thermal conductivity and a decaying oscillatory mode that determines the sound speed and the sound diffusivity. The data are well fit by a linear hydrodynamics analytic model, enabling measurement of both the shear viscosity $\eta$ and the thermal conductivity $\kappa_T$.

The experiments employ ultracold $^6$Li atoms in a balanced mixture of the two lowest hyperfine states, which are evaporatively cooled in a CO$_2$ laser trap and loaded into a box potential. The box comprises six sheets of blue-detuned light, created by two digital micromirror devices (DMDs)~\cite{LorinLinearHydro}. The top and bottom sheets employ a 669 nm beam. The four vertically propagating sheets are produced by a 532 nm beam, which passes through a diffractive  optical element and an imaging lens to produce a ``top-hat" shaped intensity profile on the surface of the DMD array. The box potential $U_0(\mathbf{r})$ yields a rectangular density profile with typical dimensions $(x,y,z)=(52\times50\times150)\,\mu$m. The density varies  slowly in the direction of the long ($z$) axis, due to the harmonic confining potential $\propto z^2$ arising from the curvature of the bias magnetic field, which has little effect on the shorter $x$ and $y$ axes.  The typical total central density is $n_0= 4.5\times10^{11}$ atoms/cm$^3$, with the Fermi energy $\epsilon_{\!F0}\equiv k_B T_F =k_B\times 0.22\,\mu$K and Fermi speed $v_F\simeq 2.5$ cm/s. The box depth $U_0\simeq 1.1\,\mu$K~\cite{SupportOnline}.

Once the cloud is loaded into the box potential, we employ the 532 nm DMD to slowly ramp up an additional small optical potential $\delta U(z)$, which is spatially periodic along the $z$-axis.  After thermal equilibrium is established, the cloud profile exhibits a periodic spatial modulation, $\delta n(z,0)$, Fig.~\ref{fig:boxdensity}. The measurements employ modulation amplitudes $\delta n/n_0$ from 7\% to 19\%,  yielding consistent data within our error bars.

After the periodic potential is abruptly extinguished, we measure the oscillatory decay of the density change $\delta n(z,t)=n(z,t)-n_0(z)$. For each image, the signal $n$ and subtracted background $n_0(z)$ densities are scaled to their respective total atom number to suppress noise arising from shot to shot atom number variation. We perform a fast Fourier transform (FFT) of $\delta n(z,t)$ at each time, in a region containing an integer number (typically 3-4) of spatial periods near the peak density, minimizing the imaginary component to obtain a real transform,  $\delta n(q,t)$, Fig.~\ref{fig:Fourier}.

To model the data, where the initial conditions are isothermal, it is convenient to construct the coupled equations for the changes in the density $\delta n(z,t)$ and temperature $\delta T(z,t)$. We use the continuity equation to eliminate the velocity field. For experiments in the linear response regime~\cite{SupportOnline},
\begin{eqnarray}
\delta\ddot{n}&=&c_T^2\,\partial_z^2(\delta n+\delta\tilde{T})
+\frac{\frac{4}{3}\,\eta+\xi_B}{n_0 m}\,\partial_z^2\delta\dot{n}\nonumber\\
& &+\,\frac{1}{m}\,\partial_z[n_0(z)\,\partial_z\delta U+\delta n\,\partial_z U_0(z)],
\label{eq:2.5}
\end{eqnarray}
with $c_T$ the isothermal sound speed and $m$ the atom mass. Here, $\delta\tilde{T}=n_0\beta\,\delta T$ has dimension of density, with $\beta=-1/n(\partial n/\partial T)_P$ the thermal expansivity~\cite{SupportOnline} and
\begin{equation}
\delta\dot{\tilde{T}}=\epsilon_{LP}\,\delta\dot{n}+
\frac{\kappa_T}{n_0\,c_{V_1}}\,\partial_z^2\delta\tilde{T},
\label{eq:2.6}
\end{equation}
with $\epsilon_{LP}\equiv c_{P_1}/c_{V_1}-1$ the Landau-Placzek parameter. The heat capacities per particle at constant volume $c_{V_1}$ and at constant pressure $c_{P_1}$ can be determined from the measured equation of state~\cite{KuThermo,SupportOnline}.

Eqs.~\ref{eq:2.5}~and~\ref{eq:2.6} have simple physical interpretations. The $c_T^2$ terms on the right-hand side of eq.~\ref{eq:2.5} correspond to the pressure change~\cite{SupportOnline}. A viscous damping force arises from the shear viscosity, $\eta\equiv\alpha_\eta\,\hbar n_0$, while the bulk viscosity $\xi_B$ vanishes for a unitary Fermi gas~\cite{Bulk}.  The final terms in eq.~\ref{eq:2.5} arise from the perturbing and box potentials, where $\partial_z U_0(z)$ is found from the slowly varying background density $n_0(z)$~\cite{SupportOnline} and $\delta U(z,t>0)=0$ for our experiments. The first term on the right-hand side of eq.~\ref{eq:2.6} describes the adiabatic change in the temperature due to the change in density. The last term describes temperature relaxation at constant density due to the heat flux, which is proportional to the thermal conductivity $\kappa_T\equiv\alpha_\kappa\,\hbar n_0\,k_B/m$. Eqs.~\ref{eq:2.5}~and~\ref{eq:2.6} can be solved numerically for $\delta n(z,t)$, with the initial conditions $\delta n(z,0)$ (measured), $\delta\dot{n}(z,0)=0$, and $\delta\tilde{T}(z,0)=0$.

We find that a perturbation wavelength $\lambda\simeq 23\,\mu$m yields good dynamic range for decay measurements over time scales that avoid perturbing $\delta n(z,t)$ in the measured central region by reflections from the walls of box potential, which then can be neglected. Since $\delta U=0$, a spatial Fourier transform of eqs.~\ref{eq:2.5}~and~\ref{eq:2.6} yields coupled equations for $\delta\ddot{n}(q,t)$ and $\delta\dot{\tilde{T}}(q,t)$. These determine the analytic fit function~\cite{SupportOnline},
\begin{equation}
\delta n(q,t)=A_0\,e^{-\Gamma t}\!+e^{-a t}\left[A_1\cos(b\,t)+A_2\,\sin(b\,t)\right],
\label{eq:FTvsTime}
\end{equation}
where  $A_1=A-A_0$ and $A_2=[(\Gamma-a)A_0+a\,A]/b$ satisfy two of the initial conditions $\delta n(q,0)=A$ and $\delta \dot{n}(q,0)=0$. The third initial condition  $\delta\ddot{n}(q,0)=-c_T^2 q^2 A$~\cite{SupportOnline} determines $A_0=A(a^2+b^2-c_T^2\, q^2)/[(\Gamma -a)^2+b^2]$.

We see that the solution consists of two independent modes, Fig.~\ref{fig:Fourier} (inset). One mode is exponentially decaying and determines the thermal diffusivity as discussed below. The other is a decaying, oscillating first sound mode, which determines the sound diffusivity. Together, the decay rates of these two distinct modes determine both the thermal conductivity and the shear viscosity.

The frequencies $\Gamma$, $a$, and $b$ in eq.~\ref{eq:FTvsTime} are related to the frequencies $\gamma_\eta\equiv 4\eta\, q^2/(3n_0 m)$, $\gamma_\kappa\equiv\kappa_T q^2/(n_0c_{V_1})$, and $c_Tq$ by~\cite{SupportOnline}
\begin{eqnarray}
\Gamma + 2\,a=\gamma_{\kappa}+\gamma_{\eta}\label{eq:frequencies1}\\
a^2+b^2+2\,a\,\Gamma=c_S^2\, q^2+\gamma_{\eta}\,\gamma_{\kappa}\label{eq:frequencies2}\\
\Gamma (\,a^2+b^2)=c_T^2\,q^2\,\gamma_{\kappa}.\label{eq:frequencies3}
\end{eqnarray}
Here, $c_S$ and $c_T$ are the adiabatic and isothermal sound speeds, which obey $c_S^2/c_T^2=c_{P_1}/c_{V_1}=1+\epsilon_{LP}(T/T_F)$.

Fitting eq.~\ref{eq:FTvsTime} to the data yields the red curve in Fig.~\ref{fig:Fourier}, with the three frequencies $c_Tq,\gamma_\eta,\gamma_\kappa$, and the amplitude $A$ as free parameters. We find that fitting the data with $A_0=0$ increases the $\chi^2$ per degree of freedom from $\simeq 1$ to $\simeq 20$, demonstrating the importance of the thermal diffusion mode, which determines the thermal conductivity in our measurements.

The reduced temperature $T/T_F=\theta(c_T/v_F)$ in eqs.~\ref{eq:frequencies1}-\ref{eq:frequencies3} is self-consistently determined from $c_T$ by the equation of state~\cite{KuThermo,SupportOnline}, with $v_F$ given for the average central density $n_0$~\cite{SupportOnline}. The fits determine the frequency $c_Tq$ within 2\%, enabling in-situ thermometry~\cite{ThetaError}.

We note that in the long wavelength (LW) limit, where $c_S\,q>>\gamma_\kappa,\gamma_\eta$, eq.~\ref{eq:frequencies2} requires $b/q\simeq c_S$, the first sound speed. Then eq.~\ref{eq:frequencies3} reduces to $\Gamma/q^2\simeq\kappa_T/(n_0c_{P_1})=D_T$, the thermal diffusivity, and eq.~\ref{eq:frequencies1} yields $2\,a/q^2\simeq \gamma_\eta/q^2+\gamma_\kappa/q^2-\Gamma/q^2= D_1$, the usual first sound diffusivity~\cite{SupportOnline,LandauFluids}. In our experiments, where $\lambda\simeq 23\,\mu$m, we find that $b/q$ is smaller than $c_S$ by 2.2\,\%, 4.3\,\% and 5.7\,\% for $T/T_F=0.28,0.46,$ and $0.63$, respectively, close to the LW limit.

Further, the LW limit requires $A_0/A=1-c_{V_1}/c_{P_1}$, which is $\simeq 0.3$ for our $T/T_F$ range and within $10\,$\% of the measured values. As a cross check, we fit the data with eq.~\ref{eq:FTvsTime}, letting both $A$ and $A_0$ be free parameters, and obtain consistent results for $A_0/A$.

We also estimate the change in the energy per particle $W_1$ that arises from the energy stored in the initial spatially periodic density profile. Assuming adiabatic compression, we find~\cite{SupportOnline},
\begin{equation}
W_1=\frac{m c_S^2}{2}\left\langle\left(\frac{\delta n}{n_0}\right)^2\right\rangle.
\label{eq:energystored}
\end{equation}
For $\delta n/n_0\simeq 0.2\cos(qz)$, we have $W_1\simeq 0.01\,m c_S^2$. As $mc_S^2=10/9\, E_1$, with $E_1$ the energy per particle~\cite{MZSound,SupportOnline}, the change in $E_1$, and hence in $\theta=T/T_F$, is negligible.

\begin{figure}[h!]
\centering
\includegraphics[width=3.15in]{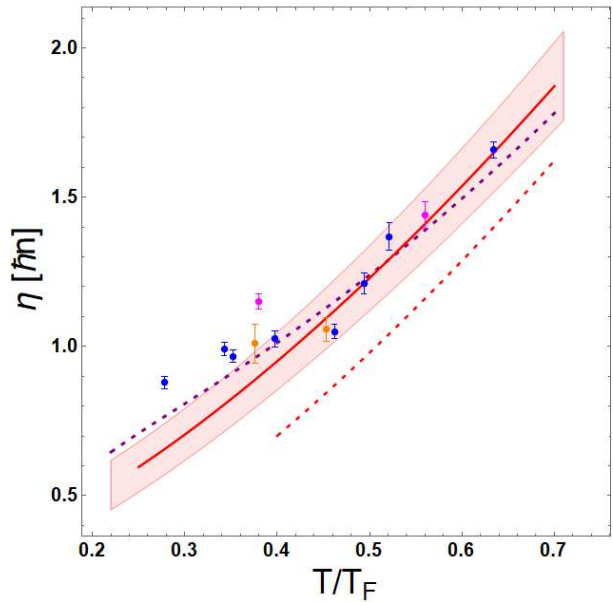}
\caption{Shear viscosity $\eta$ in units of $\hbar n$ versus reduced temperature $T/T_F$.  Blue dots: $\lambda \simeq 23\,\mu$m. Orange dots: Left\,(right) $\lambda =18.2\,(18.9)\,\mu$m.  Pink dots: Left\,(right) $\lambda = 32.3\,(41.7)\,\mu$m.  Red solid curve: Fit to cloud expansion data, $\alpha_0\,\theta^{3/2}\!\!+\alpha_2$ (Bluhm et al.,~\cite{BluhmSchaeferLocalViscosity}, see text). Shaded region denotes the standard deviation of the fit. Upper purple-dashed curve: Prediction of Enss et al.,~\cite{ZwergerViscosity}. Lower red-dashed curve: High temperature limit, $\alpha_0\,\theta^{3/2}$. Data error bars are statistical~\cite{TransportError}. (color online)
\label{fig:shearviscosity}}
\end{figure}

Our measured shear viscosity, Fig.~\ref{fig:shearviscosity}, can be compared to the high temperature diluteness expansion of Bluhm et al.,~\cite{BluhmSchaeferLocalViscosity}, $\eta_{\rm exp}(\theta)=(\alpha_0\,\theta^{3/2}+\alpha_2)\,\hbar n$, where $\alpha_0=2.77(21)$ and $\alpha_2=0.25(08)$ are measured by using a second order hydrodynamics model to fit aspect ratio data for freely expanding clouds~\cite{JosephShearNearSF}. Here, the first term is the high temperature limit, where $\theta^{3/2}n\propto T^{3/2}$ depends only on the temperature. The extracted $\alpha_0$ is in excellent agreement with a variational calculation based on the two-body Boltzmann equation for a unitary gas~\cite{BruunViscousNormalDamping,BluhmSchaeferLocalViscosity}. The leading correction from $\alpha_2$ depends only on the density. The red curve in Fig.~\ref{fig:shearviscosity} shows that $\eta_{\rm exp}(T/T_F)$  is in agreement with the measurements in the box potential for $T/T_F\geq 0.45$. For comparison, the red-dashed curve shows the high temperature limit, where $\alpha_2=0$. The top purple-dashed curve is the T-matrix theory prediction of Enss et al.,~\cite{ZwergerViscosity}, in reasonable agreement with the data.

In all of the figures, we compare data for $\lambda\simeq 23\,\mu$m to data points with $\lambda=18.2\,\mu$m, $18.9\,\mu$m, $32.3\,\mu$m (3-spatial periods) and  $41.7\,\mu$m (2-spatial periods). These  measurements show that there are no large systematic shifts with wavelength.

\begin{figure}[h!]
\centering
\includegraphics[width=3.25in]{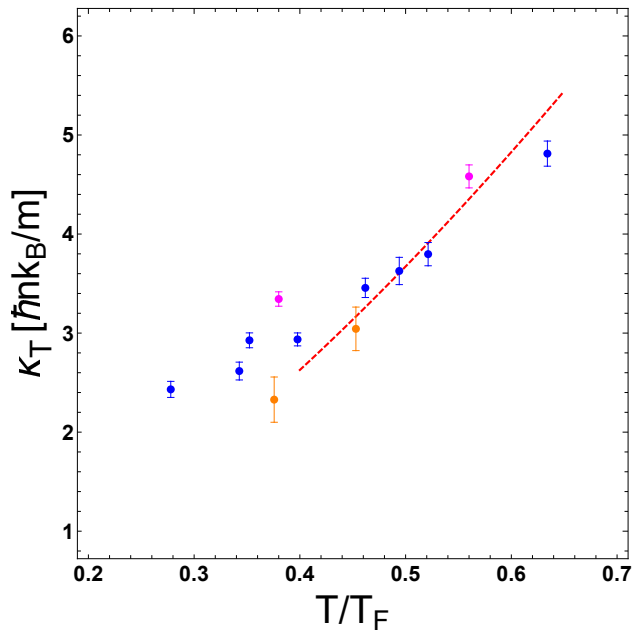}
\caption{Thermal conductivity $\kappa_T$ in units of $\hbar n\, k_B/m$ versus reduced temperature $T/T_F$. Blue dots: $\lambda \simeq 23\,\mu$m. Orange dots: Left\,(right) $\lambda =18.2\,(18.9)\,\mu$m.  Pink dots: Left\,(right) $\lambda = 32.3\,(41.7)\,\mu$m.   Red-dashed curve: High temperature limit, $15/4\,\alpha_0\,\theta^{3/2}$. Error bars are statistical~\cite{TransportError}. (color online)
\label{fig:thermalconductivity}}
\end{figure}

Our measured thermal conductivity, Fig.~\ref{fig:thermalconductivity}, can be compared with variational calculations for a unitary Fermi gas in the high temperature, two-body Boltzmann equation limit~\cite{BrabySchaeferThermalCond}, where $\kappa_T(\theta)=15/4\,\alpha_0\,\theta^{3/2}\,\hbar n\,k_B/m$, with $k_B$ the Boltzmann constant.  The red-dashed line in Fig.~\ref{fig:thermalconductivity} shows that the high temperature prediction  is in reasonable agreement with measurements in the box potential for $T/T_F\geq 0.45$, without a temperature-independent correction, but the data are significantly smaller than predicted~\cite{EnssTransport,MaThermalCond}.

\begin{figure}[h!]
\centering
\includegraphics[width=3.25in]{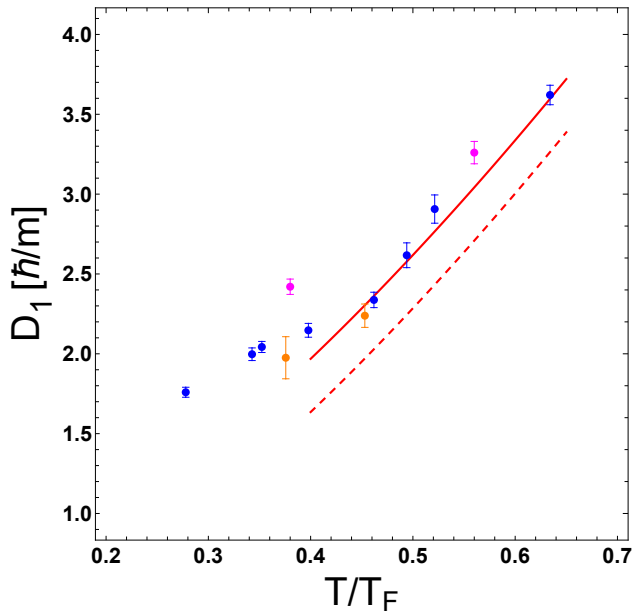}
\caption{Sound diffusivity $D_1=2\,a/q^2$, in units of $\hbar/m$ versus reduced temperature $T/T_F$.  Blue dots: $\lambda \simeq 23\,\mu$m. Orange dots: Left\,(right) $\lambda =18.2\,(18.9)\,\mu$m.  Pink dots: Left\,(right) $\lambda = 32.3\,(41.7)\,\mu$m.  Red-dashed curve: Long wavelength, high temperature limit, $D_1=7/3\, \alpha_0\,\theta^{3/2}$. Red solid curve: $D_1=4/3\,(\alpha_0\,\theta^{3/2}\!\!+\alpha_2)+ \alpha_0\,\theta^{3/2}$. Error bars are statistical~\cite{TransportError}. (color online) \label{fig:sounddiffusivity}}
\end{figure}

The sound diffusivity $D_1=2\,a/q^2$, in units of $\hbar/m$, Fig.~\ref{fig:sounddiffusivity}, is determined by eqs.~\ref{eq:frequencies1}-\ref{eq:frequencies3} from the fit parameters $c_Tq$, $\gamma_\eta$, and $\gamma_\kappa$. We obtain the same results within our error bars, by directly fitting $\Gamma$, $a$ and $b$ in eq.~\ref{eq:FTvsTime},  constraining $A_0/A$ using the long wavelength (LW) limit, where $b\simeq c_Sq$ determines $T/T_F$.  The red-dashed curve shows the predicted LW $D_1$, using the high temperature limits for both the shear viscosity and the thermal conductivity, with $c_{P_1}=5/2\,k_B$ and $c_{V_1}=3/2\,k_B$. For the red solid curve, the high temperature shear viscosity term in $D_1$ is replaced with the measured viscosity for the expanding gas, $\eta_{exp}(\theta)$, yielding a good fit for the higher temperature measurements, consistent with our measured $\eta$ and $\kappa_T$. Our diffusivity data can be compared to that of Patel et al.,~\cite{MZSound}, which is shifted upward relative to that of Fig.~\ref{fig:sounddiffusivity}, but exhibits nearly identical scaling with $T/T_F$, as discussed in the supplementary material~\cite{SupportOnline}.

In conclusion, we have independently determined the shear viscosity and thermal conductivity of a normal-fluid unitary Fermi gas in a box potential, directly from the two-mode oscillatory decay of a spatially periodic density perturbation. For the isothermal static initial conditions employed in the experiments, the thermally diffusive mode comprises $\simeq30$\% of the initial total amplitude of the dominant Fourier component, which is readily apparent in the free hydrodynamic relaxation.
This method is complementary to frequency domain techniques, where transport properties of quantum fluids have been determined by measuring the hydrodynamic linear susceptibility~\cite{HohenbergMartin,HuTwoFluidPRA,ZhangEnergyAbsSpectr}.  For reduced temperatures $T/T_F>0.45$, we find that the shear viscosity measured by free hydrodynamic relaxation in a box potential is consistent with that extracted from data on expanding clouds~\cite{BluhmSchaeferModIndep,BluhmSchaeferLocalViscosity}, which includes a significant density-dependent contribution. At lower temperatures, $T/T_F<0.4$, the shear viscosity measured in the box is consistently larger than that of the expanding cloud. The thermal conductivity for $T/T_F>0.45$ is close to the high temperature limit. In contrast to the shear viscosity, the pure density dependent contribution to the high temperature thermal conductivity appears to be quite small. These results emphasize the need for rigorous calculations of the leading density-dependent corrections to the two-body high temperature limits.  Finally, we expect that in the superfluid regime, the exponentially decaying mode will evolve into an oscillating second sound mode, which we hope to study in future experiments.

\noindent{\it Note added.}- After submission of our paper, a related study appeared ~\cite{SecondSoundLi}.

\vspace{0.5in}We thank Thomas Sch\"{a}fer for stimulating discussions and Parth Patel and Martin Zwierlein for providing their sound diffusivity data in table form. Primary support for this research is provided by the Physics Divisions of the National Science Foundation (PHY-2006234) and the Air Force Office of Scientific Research (FA9550-16-1-0378).\\

$^*$Corresponding author: jethoma7@ncsu.edu

%

\newpage
\widetext
\setcounter{figure}{0}
\setcounter{equation}{0}
\renewcommand{\thefigure}{S\arabic{figure}}
\renewcommand{\theequation}{S\arabic{equation}}

\appendix
\section{Supplemental Material}

In this supplemental material, we derive a linear hydrodynamics  model to analyze the free oscillatory decay of a spatially periodic density perturbation in a normal fluid unitary Fermi gas. An analytic fit function is found for the time-dependence of the dominant spatial Fourier component, which determines the sound speed, the shear viscosity, and the thermal conductivity. We discuss the sound diffusivity, the thermodynamics employed in the model, the energy stored in the initial periodic density profile, the determination of the 3D density, and the systematic error arising from the density variation.

\subsection{Hydrodynamic linear response for a normal fluid.}
We consider a normal fluid Fermi gas, which is a single component fluid with a mass density is $\rho\equiv n\,m$, where $n$ is the total particle density (we assume a 50-50 mixture of two components) and $m$ is the atom mass. $\rho({\mathbf{r}},t)$ satisfies the continuity equation,
\begin{equation}
\partial_t\rho +\partial_i(\rho\,v_i)=0,
\label{eq:1.3}
\end{equation}
where a sum over $i=x,y,z$ is implied. The mass flux (momentum density) is $\rho\,v_i$, with $v_i({\mathbf{r}},t)$ the velocity field.

The momentum density and corresponding momentum flux $\rho\,v_i v_j$ obey
\begin{equation}
\partial_t (\rho\,v_i) +\partial_j (\rho\,v_i v_j)=-\partial_i p -n\,\partial_i U+\partial_j(\eta\,\sigma_{ij}+\xi_B\,\sigma'_{ij}),
\label{eq:1.5}
\end{equation}
Here, $-\partial_i p-n\,\partial_i U$ is the force per unit volume arising from the pressure $p$ and the externally applied potential $U({\mathbf{r}},t)$. The last terms describe the dissipative forces, which arise generally from the shear viscosity  $\eta$ and the bulk viscosity $\xi_B$, with $\sigma_{ij}\equiv\partial_iv_j+\partial_jv_i-2\,\delta_{ij}\nabla\cdot{\mathbf{v}}/3$ and $\sigma'_{ij}=\delta_{ij}\nabla\cdot{\mathbf{v}}$.  For generality, we retain $\xi_B$, which vanishes for a unitary gas~\cite{SonBulkViscosity,StringariBulk,ElliottScaleInv}. Taking the divergence of eq.~\ref{eq:1.5}, and using eq.~\ref{eq:1.3}, we immediately obtain
\begin{equation}
-\partial_t^2\rho+\partial_i\partial_j (\rho\,v_i v_j)=-\partial_i^2 p-\partial_i(n\,\partial_iU)+\partial_i\partial_j(\eta\,\sigma_{ij}+\xi_B\,\sigma'_{ij}).
\label{eq:2.7}
\end{equation}

We are interested in the  hydrodynamic linear response to a perturbing external potential $\delta U({\mathbf{r}},t)$, which leads to first order changes in the density $\delta n({\mathbf{r}},t)$ and pressure $\delta p({\mathbf{r}},t)$,
\begin{eqnarray}
n({\mathbf{r}},t)&=&n_0({\mathbf{r}})+\delta n({\mathbf{r}},t)\nonumber\\
p({\mathbf{r}},t)&=&p_0({\mathbf{r}})+\delta p({\mathbf{r}},t)\nonumber\\
U({\mathbf{r}},t)&=&U_0({\mathbf{r}})+\delta U({\mathbf{r}},t).
\label{eq:3.1}
\end{eqnarray}
Here, $n_0({\mathbf{r}})$ and $p_0({\mathbf{r}})$ are the equilibrium (time independent) density and pressure arising from confinement in the box trap potential, $U_0({\mathbf{r}})$. In equilibrium, the velocity field ${\mathbf v}_0({\mathbf{r}},t)=0$ and eq.~\ref{eq:1.5} requires balance of the forces per unit volume arising from the box trap and the pressure,
\begin{equation}
-\nabla p_0({\mathbf{r}})-n_0({\mathbf{r}})\nabla U_0({\mathbf{r}})=0.
\label{eq:1.3b}
\end{equation}
Substituting eq.~\ref{eq:3.1} into  eq.~\ref{eq:2.7} and retaining terms to first order in small quantities, we obtain
\begin{equation}
\partial_t^2\delta n=\frac{1}{m}\nabla^2\,\delta p+\frac{1}{m}\nabla\cdot[n_0({\mathbf{r}})\,\nabla\delta U+\delta n\,\nabla U_0]-\frac{1}{m}\partial_i\partial_j(\eta\,\sigma_{ij}+\xi_B\,\sigma'_{ij}).
\label{eq:3.7}
\end{equation}
Here, the second term on the left side of eq.~\ref{eq:2.7} is negligible, as the velocity field is first order in small quantities.

To evaluate the last term in eq.~\ref{eq:3.7}, we assume that the dissipative forces are small compared to the conservative forces and that the density $n_0$ slowly varies in the region of interest. Then we can ignore the spatial derivatives of $\eta$, $\xi_B$, and $n_0$, yielding
$$\partial_i\partial_j(\eta\,\sigma_{ij}+\xi_B\,\sigma'_{ij})\simeq\eta\,\partial_i\partial_j\sigma_{ij}
+\xi_B\,\partial_i\partial_j\sigma'_{ij}\simeq \left(\frac{4}{3}\,\eta+\xi_B\right)\,\nabla^2(\nabla\cdot{\mathbf{v}}).$$
The velocity field is eliminated using $\nabla\cdot{\mathbf{v}}\simeq -\partial_t\delta n/n_0=-\delta\dot{n}/n_0$, which follows from  eq.~\ref{eq:1.3}.
With eq.~\ref{eq:deltap} of \S~\ref{sec:thermo} $\delta p=mc_T^2(\delta n+\delta\tilde{T})$, eq.~\ref{eq:3.7} becomes
\begin{eqnarray}
\delta\ddot{n}=c_T^2\,\nabla^2(\delta n+\delta\tilde{T})
+\frac{1}{m}\nabla\cdot[n_0({\mathbf{r}})\,\nabla\delta U+\delta n\,\nabla U_0]+\frac{\frac{4}{3}\eta+\xi_B}{n_0 m}\,\nabla^2\delta\dot{n},
\label{eq:2.5S}
\end{eqnarray}
where $\delta\tilde{T}\equiv\beta\, n\,\delta{T}$, from eq.~\ref{eq:1.2Sb}.

To complete the model, we require the evolution equation for $\delta\tilde{T}$, which is determined from eq.~\ref{eq:1.7Sc} of \S~\ref{sec:thermo},
\begin{equation}
\delta\dot{T}=\epsilon_{LP}\,\frac{\delta\dot{n}}{\beta\,n}+\frac{T\delta\dot{s}_1}{c_{V_1}}.
\label{eq:1.9S}
\end{equation}
Here, $\epsilon_{LP}\equiv c_{P_1}/c_{V_1}-1$ the Landau-Placzek parameter, $T = T_0$ is the initial, spatially-uniform, temperature and $n=n_0$ is the initial spatially-uniform density.

The heating rate per particle, $T\delta\dot{s}_1$ is determined to first order in small quantities by
\begin{equation}
T(\partial_t+{\mathbf v}\cdot\nabla)\,\delta s_1=T\,\delta\dot{s}_1=\frac{\delta\dot{q}}{n_0},
\label{eq:4.2}
\end{equation}
where $\delta\dot{q}$ the heating rate per unit volume. The heating rate arising from the shear viscosity is second order in $v_i$, which is negligible compared to the heating rate arising from heat conduction. Hence, $\delta\dot{q}\simeq -\nabla\cdot(-\kappa_T\nabla\delta T)\simeq\kappa_T\nabla^2\delta T$, where we neglect the spatial derivatives of $\kappa_T$ and
\begin{equation}
\delta\dot{T}=\epsilon_{LP}\,\frac{\delta\dot{n}}{\beta\,n}+\frac{\kappa_T}{n c_{V1}}\nabla^2\delta T.
\label{eq:1.9Sb}
\end{equation}
Multiplying eq.~\ref{eq:1.9Sb} by $\beta\, n$, we obtain finally
\begin{equation}
\delta\dot{\tilde{T}}=\epsilon_{LP}\,\delta\dot{n}+\frac{\kappa_T}{n_0 c_{V1}}\nabla^2\delta\tilde{T}.
\label{eq:1.9Sc}
\end{equation}
Eqs.~\ref{eq:2.5S}~and~\ref{eq:1.9Sc} determine the evolution for the given forces $\nabla U_0$ and $\nabla\delta U$.

For the experiments, we employ a one-dimensional approximation, where $\partial_z U_0(z)$ can be determined from the background density profile $n_0(z)$, as shown in \S~\ref{subsec:boxforce}, and $\delta U=0$. Eqs.~\ref{eq:2.5S}~and~\ref{eq:1.9Sc} are numerically integrated, subject to three initial conditions. In the experiments, $\delta n(z,0)$,  is measured by imaging the cloud. Static equilibrium requires $\delta \dot{n}(z,0)=0$ and the isothermal condition requires $\delta\tilde{T}(z,0)=0$.

A fast Fourier transform of the predicted density perturbation $\delta n(z,t)$,  yields the Fourier component $\delta n(q,t)$, which is evaluated for the wave vector $q$ corresponding to the peak of the Fourier transform. The predicted time-dependent $\delta n(q,t)$, obtained by numerical integration of eqs.~\ref{eq:2.5S}~and~\ref{eq:1.9Sc} can be used in a $\chi^2$  fit to the measured density profiles at each time, to extract the transport coefficients. Further, the numerical method determines the time scale over which the box potential has a negligible effect on the spatial region of interest, enabling the determination of an analytic fit function for $\delta n(q,t)$, which we use to fit the data.

\subsubsection{Analytic Fit Function}
\label{sec:Analytic}

In practice, it is convenient to limit the spatial region for the Fourier transform to the region near the center of the box, where the background density slowly varies. Further, the phase of the transform is selected so that the Fourier amplitudes are real, by choosing an integral number of periods for the length of the transformed region.  When the evolution is measured over short enough time scales, the box potential makes a negligible contribution to the time-dependent density profile in the region of interest. Ignoring the box potential, and noting that the evolution is measured after extinguishing $\delta U$, a spatial Fourier transform of eqs.~\ref{eq:2.5S}~and~\ref{eq:1.9Sc} yields coupled time-dependent equations for the Fourier amplitudes $\delta{n}(q,t)$ and $\delta\tilde{T}(q,t)$,
\begin{equation}
\delta\ddot{n}(q,t)=-\,\omega_T^2\, [\,\delta n(q,t)+\delta\tilde{T}(q,t)\,]-\gamma_{\eta}(q)\,\delta\dot{n}(q,t)
\label{eq:1.8FT}
\end{equation}
\begin{equation}
\delta\dot{\tilde{T}}(q,t)=\epsilon_{LP}\,\delta\dot{n}(q,t)-\gamma_{\kappa}(q)\,\delta\tilde{T}(q,t),
\label{eq:1.9FT}
\end{equation}
with $\epsilon_{LP}=c_{P_1}/c_{V_1}-1$ and $\omega_T=c_T\, q$.  Here, $\gamma_\kappa(q)\equiv\kappa_T q^2/(n_0c_{V_1})$ and $\gamma_\eta(q)\equiv 4\eta\, q^2/(3n_0 m)$, where the bulk viscosity $\xi_B=0$ in eq.~\ref{eq:2.5S}  for a unitary Fermi gas.
Assuming static initial conditions, we solve eqs.~\ref{eq:1.8FT}~and~\ref{eq:1.9FT} assuming $\delta n(q,0)\neq 0$ (measured), $\delta\dot{n}(q,0)=0$ and $\delta\tilde{T}(q,0)=0$.

We can express the local shear viscosity in units of $\hbar\,n_0$,
\begin{equation}
\eta_0\equiv \alpha_{\eta}\,\hbar\,n_0
\label{eq:alphashear}
\end{equation}
and determine $\alpha_{\eta}$ from the measurements.
Similarly, we express the thermal conductivity in units of  $\hbar\,n_0\,k_B/m$ as
\begin{equation}
\kappa_T\equiv \alpha_{\kappa}\hbar\,n_0\,\frac{k_B}{m}
\label{eq:alphakappa}
\end{equation}
and determine $\alpha_{\kappa}$ from the measurements.

Then, in eqs.~\ref{eq:1.8FT}~and~\ref{eq:1.9FT}, the $q$-dependent frequencies are
\begin{eqnarray}
\omega_T&=&c_T \,q\nonumber\\
\gamma(q)&=&\frac{\hbar}{m}\,q^2\nonumber\\
\gamma_{\eta}(q)&=&\frac{4}{3}\,\alpha_{\eta}\,\gamma(q)\nonumber\\
\gamma_{\kappa}(q)&=&\alpha_{\kappa}\,\frac{k_B}{c_{V_1}}\gamma(q).
\label{eq:frequencies}
\end{eqnarray}

We obtain an analytic solution to eqs.~\ref{eq:1.8FT}~and~\ref{eq:1.9FT}, by assuming modes of the form $\delta n(q,t)=A e^{-s\, t}$ and $\delta \tilde{T}(q,t)=B e^{-s\, t}$, which requires
\begin{eqnarray}
(s^2+\omega_T^2-\gamma_\eta\, s)\,A+\omega_T^2\, B&=&0\nonumber\\
\epsilon_{LP}\, s\, A-(s-\gamma_\kappa)\,B&=&0.
\label{eq:1.4AS}
\end{eqnarray}
A nontrivial solution is obtained by setting the determinant of the coefficients equal to zero,
\begin{equation}
s^3-s^2\,(\gamma_\kappa+\gamma_\eta)+s\,(\omega_S^2+\gamma_\kappa\gamma_\eta)-\omega_T^2\gamma_\kappa=0,
\label{eq:1.7AS}
\end{equation}
where $\omega_S^2=(1+\epsilon_{LP})\,\omega_T^2=c_{P_1}/c_{V_1}\,\omega_T^2$, i.e., $\omega_T=c_T \,q$ and $\omega_S=c_S \,q$.

Eq.~\ref{eq:1.7AS} is a cubic polynomial with real coefficients, which must have one real root and one complex pair, i.e., it factors as $(s-\Gamma)\,[(s-a)^2+b^2]$. Then,
\begin{equation}
s^3-s^2\,(\Gamma + 2\,a)+s\,(a^2+b^2+2\,a\,\Gamma)-\Gamma\,(a^2+b^2)=0.
\label{eq:1.11AS}
\end{equation}
Comparing the coefficients of $s^n$ in eq.~\ref{eq:1.11AS} and eq.~\ref{eq:1.7AS}, we find
\begin{eqnarray}
\Gamma + 2\,a=\gamma_{\kappa}+\gamma_{\eta}\nonumber\\
a^2+b^2+2\,a\,\Gamma=c_S^2\, q^2+\gamma_{\eta}\,\gamma_{\kappa}\nonumber\\
\Gamma (\,a^2+b^2)=c_T^2\,q^2\,\gamma_{\kappa},
\label{eq:frequenciesAS}
\end{eqnarray}

As there are three solutions with three initial conditions, we take the density perturbation to be
\begin{equation}
\delta n(q,t)=A_0\,e^{-\Gamma t}\!+e^{-a t}\left[A_1\cos(b\,t)+A_2\,\sin(b\,t)\right],
\label{eq:2.9AS}
\end{equation}
where  $A_1=A-A_0$ and $A_2=[(\Gamma-a)A_0+a\,A]/b$ satisfy two of the initial conditions $\delta n(q,0)=A$ and $\delta\dot{n}(q,0)=0$.
With $\delta\tilde{T}(q,0)=0$, the third initial condition follows from eq.~\ref{eq:1.8FT}, $\delta\ddot{n}(q,0)=-\omega_T^2\,A$. Using eq.~\ref{eq:2.9AS}, this yields the amplitude $A_0$,
\begin{equation}
[(\Gamma -a)^2+b^2]A_0 = (a^2+b^2-c_T^2 q^2)A.
\label{eq:3.8A0}
\end{equation}

Similarly, the temperature perturbation is given by
\begin{equation}
\delta\tilde{T}(q,t)=B_0\,\left[e^{-\Gamma t}\!-e^{-a t}\cos(b\,t)+\frac{\Gamma -a}{b}\,e^{-a t}\sin(b\,t)\right],
\label{eq:7.2AS}
\end{equation}
which satisfies $\delta\tilde{T}(q,0)=0$ and $\delta\dot{\tilde{T}}(q,0)=0$, as required by eq.~\ref{eq:1.9FT} with the initial condition $\delta\dot{n}(q,0)=0$. From eq.~\ref{eq:1.9FT}, we also have the additional constraint $\delta\ddot{\tilde{T}}(q,0)=-\epsilon_{LP}\,c_T^2q^2\,A$. Using eq.~\ref{eq:7.2AS}, we find $B_0$,
\begin{equation}
[(\Gamma -a)^2+b^2]B_0 = -\epsilon_{LP}\,c_T^2 q^2\,A.
\label{eq:7.9AS}
\end{equation}

As described in the main text, we fit eq.~\ref{eq:2.9AS} to the data using the three frequencies $c_Tq,\gamma_\eta,\gamma_\kappa$, and the amplitude $A$ as free parameters. In eqs.~\ref{eq:frequenciesAS}, note that $c_S^2/c_T^2=1+\epsilon_{LP}(T/T_F)$. Here, the reduced temperature $T/T_F=\theta(c_T/v_F)$ is self-consistently determined from $c_T=\omega_T/q$ by the equation of state~\cite{KuThermo}, see Fig.~\ref{fig:soundspeed}. The fits determine $\omega_T$ within 2\%, enabling in-situ thermometry. The primary uncertainty in $T/T_F$ arises from the uncertainty in $v_F$, which is determined by the measured density, see \S~\ref{sec:CentralDensity}.

\begin{figure}[h!]
\centering
\includegraphics[width=3.0in]{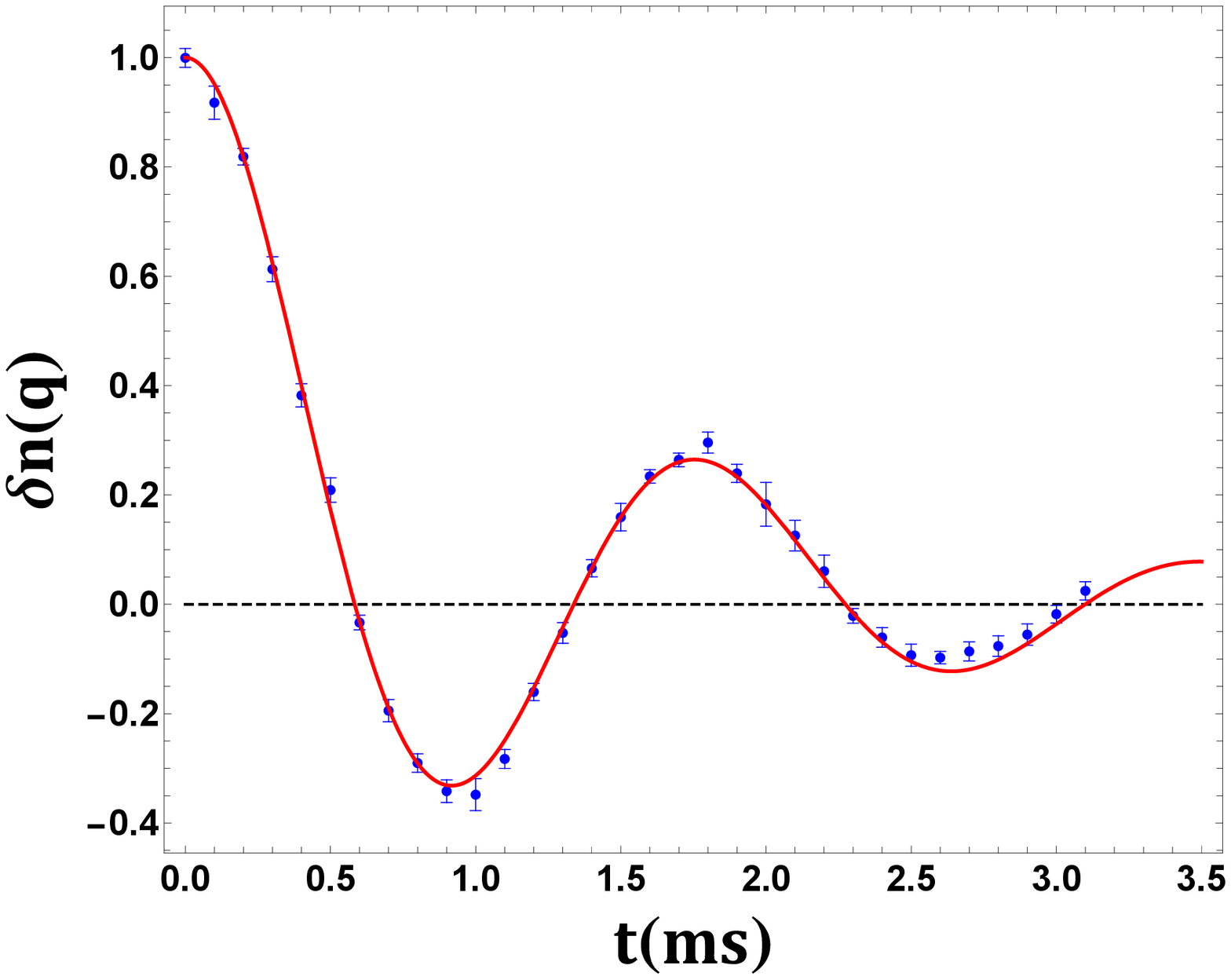}\hspace{0.25in}\includegraphics[width=3.0in]{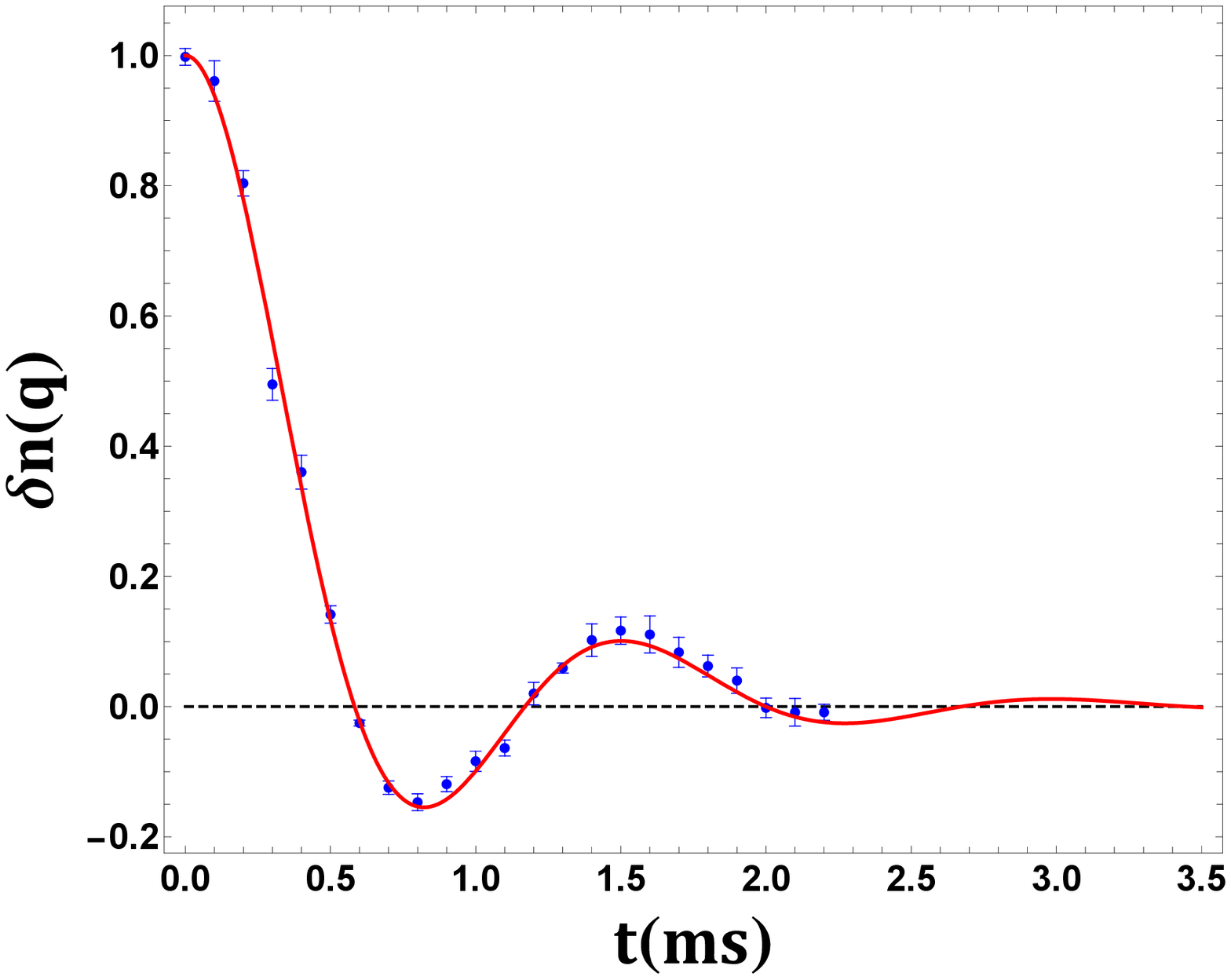}
\caption{Fit of eq.~\ref{eq:2.9AS} to the real part of the Fourier transform of the measured density for $q=2 \pi/\lambda$. Left: $T/T_F =0.28$ with $\lambda = 23.5\,\mu$m. Right: $T/T_F=0.63$ with $\lambda = 23.3\,\mu$m. Blue dots (data); Red curve (hydrodynamic model).
\label{fig:Fit}}
\end{figure}

\begin{figure}[h!]
\centering
\includegraphics[width=3.0in]{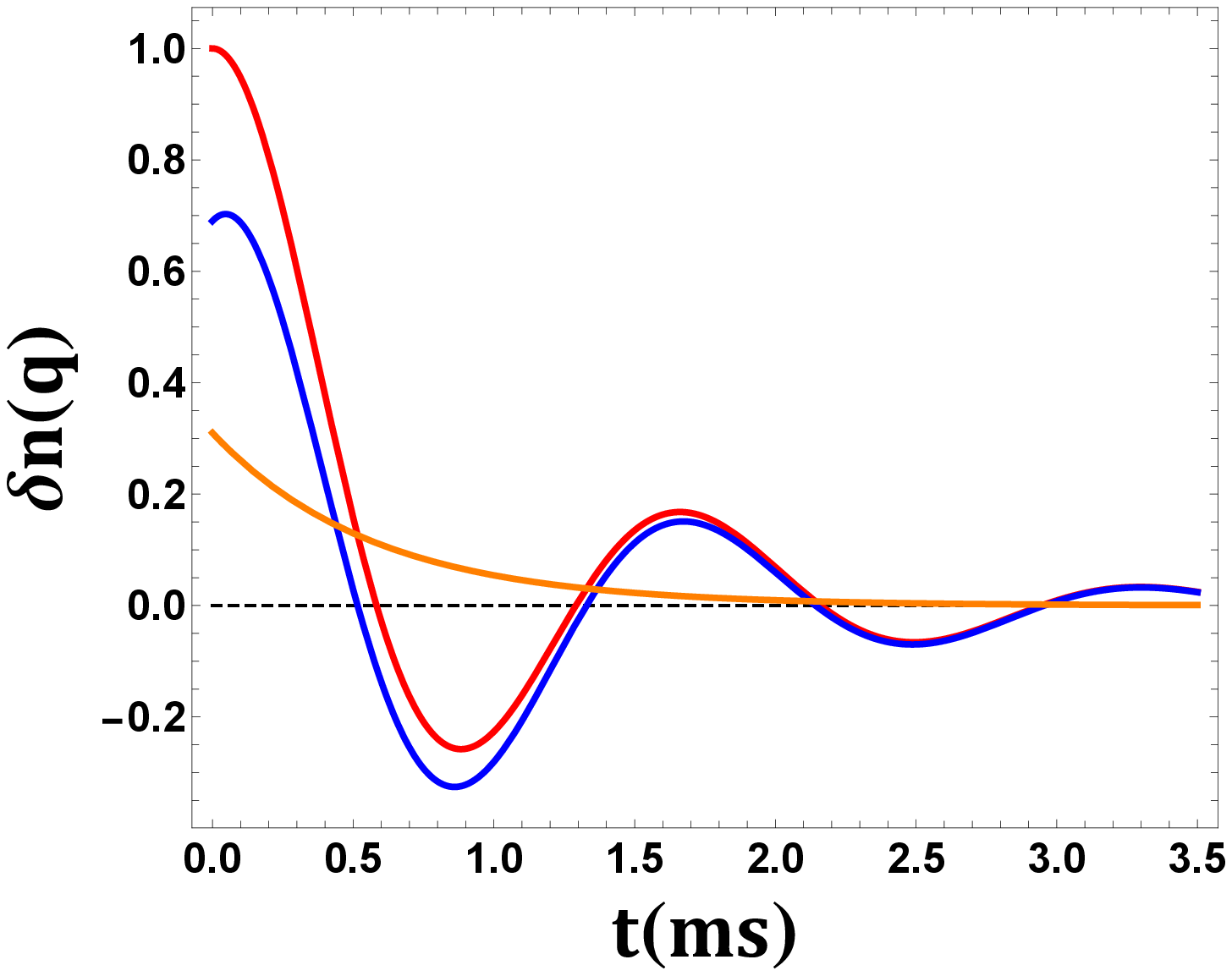}\hspace{0.25in}\includegraphics[width=3.05in]{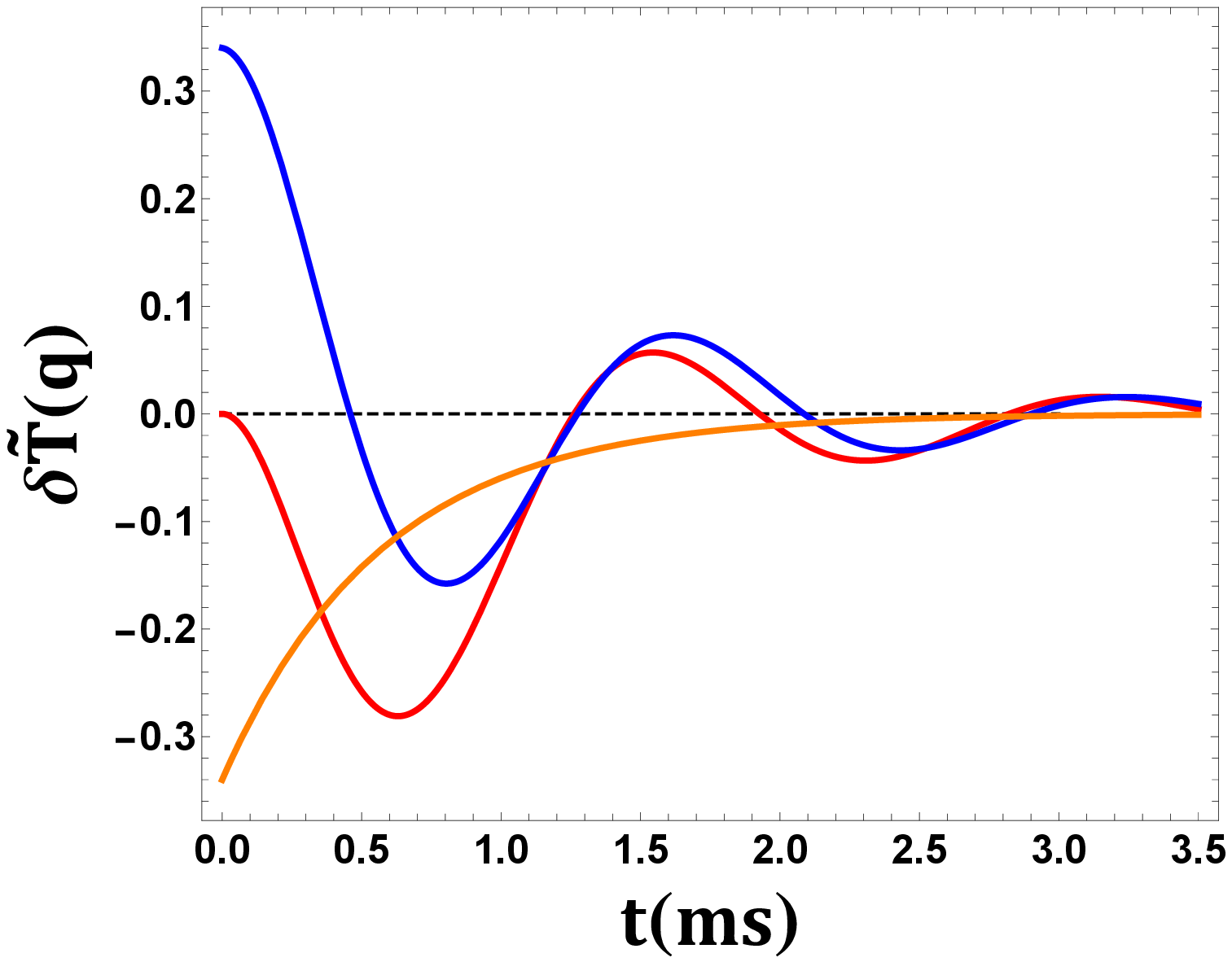}
\caption{Components of the fit function $\delta n(q,t)$ eq.~\ref{eq:2.9AS} for Fig.~2 of the main text, where  $T/T_F=0.46$ and $q=2\pi/\lambda$, with $\lambda=22.7\,\mu$m. Red curve: Total fit function; Orange curve: Zero frequency, exponentially decaying (thermal diffusion) mode; Blue curve: oscillating, exponentially decaying first sound mode. $\delta\tilde{T}(q,t)$  shows the corresponding components for the temperature perturbation (red curve).
\label{fig:Fit046Comp}}
\end{figure}

Typical fits of eq.~\ref{eq:2.9AS} for $T/T_F =0.28$ and $0.63$ are shown as the red curves in Fig.~\ref{fig:Fit}, where both the data and the model have been divided by the fit amplitude $A$.
From the fits, we can find the contributions of the first sound mode and thermal diffusion mode to $\delta n(q,t)$ and $\delta\tilde{T}(q,t)$.
First, we determine the frequencies $\Gamma$, $a$, and $b$ from the fit parameters $c_Tq,\gamma_\eta,\gamma_\kappa$ using eqs.~\ref{eq:frequenciesAS}. This is most easily done by finding the real solution $\Gamma$ of eq.~\ref{eq:1.7AS}. Then the first of eqs.~\ref{eq:frequenciesAS} determines $a$ and the second yields $b$.
Eq.~\ref{eq:3.8A0} then determines $A_0$ in terms of the fitted amplitude $A$ and eq.~\ref{eq:7.9AS} determines $B_0$. Results for Fig.~2 of the main text, where $T/T_F=0.46$ are shown in Fig.~\ref{fig:Fit046Comp}. The figure shows that the contribution of the thermal diffusion mode to $\delta n(q,0)$ is initially $\simeq 30$\%. The large amplitude enables independent determination of the thermal conductivity through the decay rate $\Gamma$. In $\delta\tilde{T}(q,t)$, we see that $\delta\tilde{T}(q,0)=0$ forces the first sound and thermal diffusion modes to be initially out of phase.

It is interesting to consider the long wavelength limit, where $c_S\,q>>\gamma_\kappa,\gamma_\eta$, although we do not require this approximation in the initial analysis of our data. In this case, the second of eqs.~\ref{eq:frequenciesAS} requires $b/q\simeq c_S$, which is the first sound speed. Recalling that $\gamma_\kappa=\kappa_T q^2/(n_0c_{V_1})$, $\gamma_\eta= 4\eta\, q^2/(3n_0 m)$, and $c_T^2/c_S^2=c_{V_1}/c_{P_1}$, we see that the last of eqs.~\ref{eq:frequenciesAS} yields $\Gamma/q^2\simeq c_T^2/c_S^2\,\gamma_\kappa/q^2=\kappa_T/(n_0c_{P_1})=D_T$, which is the thermal diffusivity. The first of eqs.~\ref{eq:frequenciesAS} then gives $2\,a/q^2\simeq \gamma_\eta/q^2+\gamma_\kappa/q^2-\Gamma/q^2= D_1$, which is the usual the first sound diffusivity~\cite{LandauFluids},
\begin{equation}
 D_1=\frac{4}{3}\frac{\eta}{n_0 m}+\left(\frac{1}{c_{V_1}}-\frac{1}{c_{P_1}}\right)\,\frac{\kappa_T}{n_0}.
 \label{eq:D1}
 \end{equation}
In the same limit, eq.~\ref{eq:3.8A0}  requires $A_0/A=1-c_{V_1}/c_{P_1}$ and eq.~\ref{eq:7.9AS} requires $B_0=-A_0$.

Our fits allow an estimate of the deviation from the long wavelength (LW) limit, which holds when products of the decay rates in eq.~\ref{eq:frequenciesAS} are sufficiently small compared to the square of the sound frequencies $c_Tq$ and $c_Sq$. For each $T/T_F$, we find that $A_0/A$, calculated from the fit parameters using eq.~\ref{eq:3.8A0}, is within 10\% of the long wavelength limit, $A_0/A=1-c_{V_1}/c_{P_1}\simeq0.3$. Further, using eqs.~\ref{eq:frequenciesAS} to find the frequency $b$ from the fit parameters ( $\omega_T$, $\gamma_\kappa$, and $\gamma_\eta$), we compute the deviation $(\omega_S-b)/\omega_S$, which we find  to be 2.2\% for $T/T_F=0.28$, 4.3\% for $T/T_F=0.46$, and a maximum of 5.7\% for $T/T_F=0.63$. These results show that the experiments are performed close to the LW regime.

\subsubsection{Sound Diffusivity}
\begin{figure}[h!]
\centering
\includegraphics[width=3.5in]{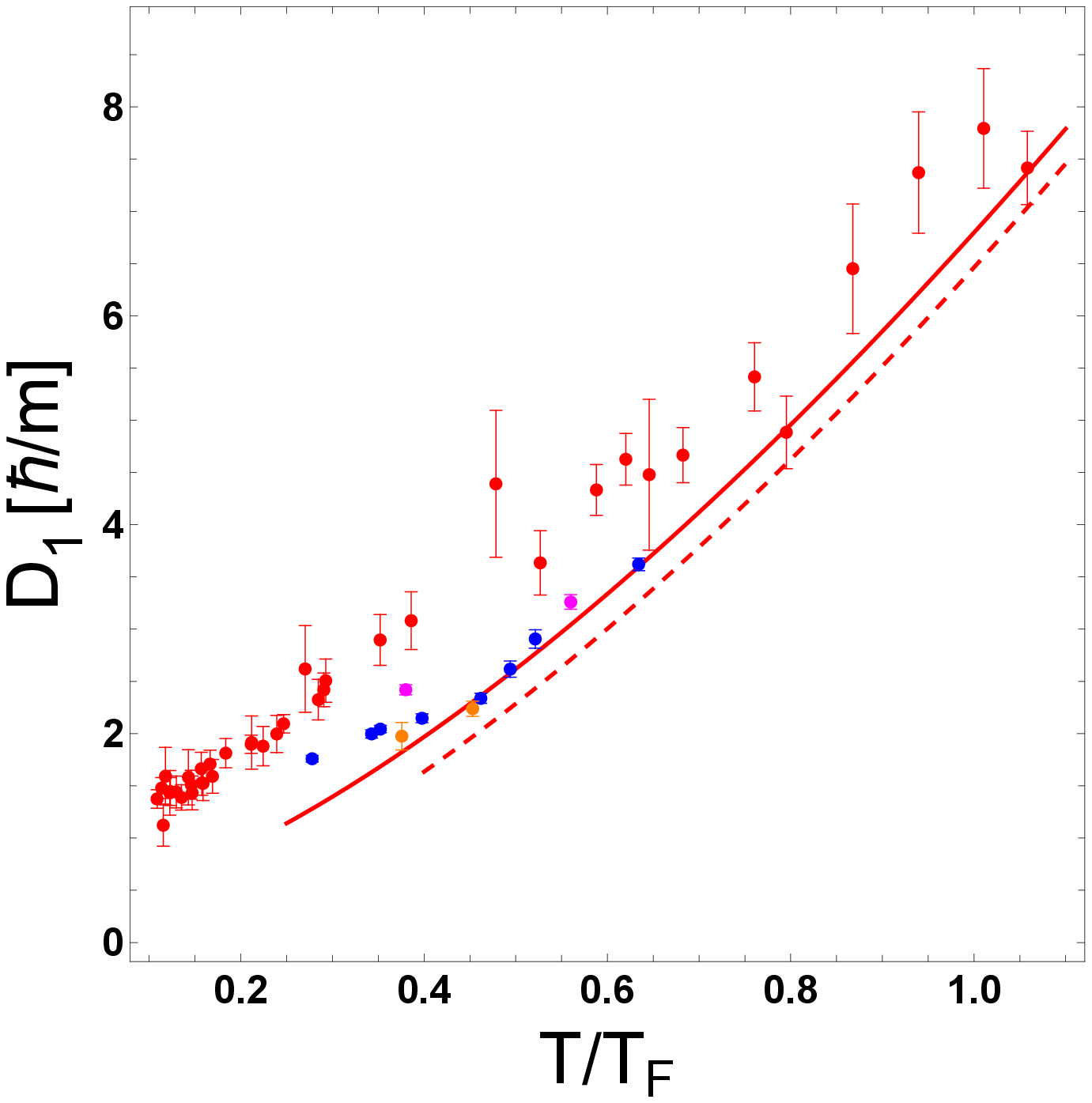}
\caption{Sound Diffusivity. $D_1=2\,a/q^2$, in units of $\hbar/m$ versus reduced temperature $\theta=T/T_F$. Red dots: Patel et al.,~\cite{MZSound}. Blue, Orange and Pink dots: Current work (see main text). Red-dashed curve: Long wavelength, high temperature limit, $D_1=7/3\, \alpha_0\,\theta^{3/2}$. Red solid curve: $D_1=4/3\,(\alpha_0\,\theta^{3/2}\!\!+\alpha_2)+ \alpha_0\,\theta^{3/2}$. \label{fig:SoundDiff}}
\end{figure}

The fits of eq.~\ref{eq:2.9AS} to the data, as in Fig.~\ref{fig:Fit}, determine the sound diffusivity $2\,a/q^2$ shown in Fig.~\ref{fig:SoundDiff}. Here, $2\,a/q^2$ takes the form of eq.~\ref{eq:D1} in the long wavelength limit.   As discussed in the main paper, the red-dashed and solid-red curves in Fig.~\ref{fig:SoundDiff} are based on the high-temperature limit of $D_1$, eq.~\ref{eq:D1}, where $1/c_{V_1}-1/c_{P_1}=4 /(15\,k_B)$  and  $\kappa_T=15\,k_B/(4\,m)\,\eta$.  The red-dashed curve shows the result for the extreme high-temperature limit $\eta=\alpha_0\,\theta^{3/2}\,\hbar\,n_0$, where $\alpha_0=2.77$~\cite{BruunViscousNormalDamping,BluhmSchaeferLocalViscosity}. The red-solid curve shows the result obtained using the diluteness expansion for the viscosity term $\eta_{\rm exp}(\theta)=(\alpha_0\,\theta^{3/2}+\alpha_2)\,\hbar n_0$, which contains a temperature independent correction~\cite{BluhmSchaeferLocalViscosity} $\alpha_2=0.25$ that is consistent with our shear viscosity measurements. However, we retain the extreme high-temperature limit for the contribution of the thermal conductivity, which appears to have a smaller temperature independent contribution. These results are discussed in the main text. The good fit to $D_1$ demonstrates the consistency of our extracted transport properties.

Fig.~\ref{fig:SoundDiff} compares our data with those of Ref.\cite{MZSound} (red dots). In both data sets, the error bars are statistical, and denote $1\,\sigma$. For our data, we show the statistical error from the $\chi^2$ fits, as described in the main text. We estimate a systematic {\it downward} shift of $\leq 5$\%,  arising from the density variation, as discussed in \S~\ref{sec:density}. For Ref.\cite{MZSound}, the estimated systematic error is 13\%,  arising from the width of the end caps.  We observe an upward shift of the diffusivity data of Ref.\cite{MZSound}, compared to that of the present work, but the scaling of the normal fluid diffusivity with reduced temperature $T/T_F$ is in good agreement.

\subsection{Thermodynamics}

\subsubsection{Thermodynamic relations}
\label{sec:thermo}

For completeness, we derive the elementary thermodynamic relations that appear in our hydrodynamic model.
Defining the density $n=1/V_1$ in terms of the volume per particle $V_1$, the expansivity is
\begin{equation}
\beta\equiv\frac{1}{V_1}\left(\frac{\partial V_1}{\partial T}\right)_p=-\frac{1}{n}\left(\frac{\partial n}{\partial T}\right)_p,
\label{eq:beta}
\end{equation}
which has a dimension of inverse temperature.

The isothermal sound speed $c_T$ is defined by
\begin{equation}
m\,c_T^2=\left(\frac{\partial p}{\partial n}\right)_{\!T}=-\left(\frac{\partial p}{\partial T}\right)_{\!\!n}\!\!\left(\frac{\partial T}{\partial n}\right)_{\!\!p},
\label{eq:cT}
\end{equation}
where we have used the chain rule. Similarly, the adiabatic sound speed $c_S$ is defined by
\begin{equation}
m\,c_S^2=\left(\frac{\partial p}{\partial n}\right)_{\!\!s_1}=-\left(\frac{\partial p}{\partial s_1}\right)_{\!\!n}\!\!\left(\frac{\partial s_1}{\partial n}\right)_{\!\!p},
\label{eq:cS}
\end{equation}
where we have defined $s_1$ as the entropy per particle.

Taking the ratio of eqs.~\ref{eq:cT}~and~\ref{eq:cS}, and using $(\partial T/\partial n)_p=1/(\partial n/\partial T)_p$
and $1/(\partial p/\partial s_1)_n=(\partial s_1/\partial p)_n$, we  obtain the well-known relation
\begin{equation}
\frac{c_T^2}{c_S^2}=\frac{\left(\frac{\partial s_1}{\partial p}\right)_{\!n}\!\left(\frac{\partial p}{\partial T}\right)_{\!n}}{\left(\frac{\partial s_1}{\partial n}\right)_{\!p}\left(\frac{\partial n}{\partial T}\right)_{\!p}}
=\frac{\left(\frac{\partial s_1}{\partial T}\right)_{\!n}}{\left(\frac{\partial s_1}{\partial T}\right)_{\!p}}=\frac{c_{V_1}}{c_{P_1}},
\label{eq:1.3S}
\end{equation}
where $c_{V_1}=T(\partial s_1/\partial T)_n$ and $c_{P_1}=T(\partial s_1/\partial T)_p$ are the heat capacities per particle at constant volume and at constant pressure, respectively.

Next, we find the first order pressure change, $\delta p$, which is needed in eq.~\ref{eq:3.7}.  We have
\begin{equation}
\delta p=\left(\frac{\partial p}{\partial n}\right)_{\!\!T}\!\!\delta n+\left(\frac{\partial p}{\partial T}\right)_{\!\!n}\!\!\delta T=
\left(\frac{\partial p}{\partial n}\right)_{\!\!T}\!\left[\delta n+\left(\frac{\partial n}{\partial p}\right)_{\!\!T}\!\!\left(\frac{\partial p}{\partial T}\right)_{\!\!n}\!\!\delta T\right].
\label{eq:1.1a}
\end{equation}
The chain rule gives
\begin{equation}
\left(\frac{\partial n}{\partial p}\right)_T\!\!\left(\frac{\partial p}{\partial T}\right)_n=-\left(\frac{\partial n}{\partial T}\right)_p=\beta n,
\label{eq:1.1b}
\end{equation}
where we have used eq.~\ref{eq:beta} for the expansivity $\beta$. With eqs.~\ref{eq:1.1a}~and~\ref{eq:cT},
\begin{equation}
\delta p=m c_T^2\, (\,\delta n+\delta\tilde{T}\,),
\label{eq:deltap}
\end{equation}
where we have defined
\begin{equation}
\delta\tilde{T}\equiv\beta\, n\,\delta{T},
\label{eq:1.2Sb}
\end{equation}
which has a dimension of density.

For the first order temperature change, we have
\begin{equation}
\delta T=\left(\frac{\partial T}{\partial n}\right)_{\!\!s_1}\!\!\delta n+\left(\frac{\partial T}{\partial s_1}\right)_{\!\!n}\!\!\delta s_1=
\left(\frac{\partial T}{\partial s_1}\right)_{\!\!n}\!\left[\left(\frac{\partial s_1}{\partial T}\right)_{\!\!n}\!\!\left(\frac{\partial T}{\partial n}\right)_{\!\!s_1}\!\!\delta n+\delta s_1\right].
\label{eq:1.6S}
\end{equation}
The chain rule gives
\begin{equation}
\left(\frac{\partial s_1}{\partial T}\right)_{\!\!n}\!\!\left(\frac{\partial T}{\partial n}\right)_{\!\!s_1}=-\left(\frac{\partial s_1}{\partial n}\right)_{\!\!T},
\label{eq:1.6Sb}
\end{equation}
which we evaluate as follows. Consider $s_1[T,n(T,p)]$. Then,
\begin{equation}
c_{P_1}=T\!\left(\frac{\partial s_1}{\partial T}\right)_{\!\!p}=T\left(\frac{\partial s_1}{\partial T}\right)_{\!\!n}+
T\!\left(\frac{\partial s_1}{\partial n}\right)_{\!\!T}\!\!\left(\frac{\partial n}{\partial T}\right)_{\!\!p}=c_{V_1}-\beta\, n\, T\!\left(\frac{\partial s_1}{\partial n}\right)_{\!\!T},
\label{eq:1.7S}
\end{equation}
where we have used eq.~\ref{eq:beta}. Hence,
\begin{equation}
\left(\frac{\partial s_1}{\partial n}\right)_{\!\!T}=-\frac{c_{P1}-c_{V1}}{\beta\, n\, T}.
\label{eq:1.7Sb}
\end{equation}
With  $(\partial T/\partial s_1)_n=T/c_{V_1}$ and eq.~\ref{eq:1.6Sb}, eq.~\ref{eq:1.6S} takes the simple form,
\begin{equation}
\delta T=\left(\frac{c_{P_1}}{c_{V_1}}-1\right)\frac{\delta n}{\beta\,n}+\frac{T\delta s_1}{c_{V_1}}.
\label{eq:1.7Sc}
\end{equation}
Here, the first term is the adiabatic change in the temperature arising from the change in density. For a monatomic gas in the high temperature limit, eq.~\ref{eq:beta} with $n=p/(k_B T)$ gives $\beta\rightarrow 1/T$ and $c_{P1}/c_{V1}-1\rightarrow2/3$. Then, $\delta T/T=2/3\,\delta n/n$, i.e., $T/T_0= (n/n_0)^{2/3}$ as expected. For a unitary Fermi gas, where $s_1=k_Bf_S(\theta)$, this result holds at all temperatures, since $(\partial T/\partial n)_{s_1}=(\partial T/\partial n)_{\theta}$, with $T=\theta\,T_F$, and $T_F\propto n^{2/3}$. The second term is the temperature change arising from the heat flow per particle, $T\delta s_1=\delta q_1$.

\subsubsection{Stored Energy}

The initial density perturbation stores energy, which is converted into kinetic energy after the perturbation is extinguished and finally into heat. To show that the change in the average energy per particle is negligible, we determined the stored energy $W$ for the ideal case of an adiabatic change of the density, $\delta n$, starting from a uniform density $n_0$. As the total number of atoms does not change during the compression, we must have
\begin{equation}
\int d^3{\mathbf r}\,\delta n({\mathbf r}) =0.
\label{eq:deltan1}
\end{equation}

Now consider a small volume $\Delta V$ of the cloud, containing a small number of atoms $\Delta N = n\,\Delta V$. Changing the volume for fixed $\Delta N$, we have $d\,\Delta N = dn\,\Delta V+n\,d\,\Delta V=0$. Taking $n\simeq n_0$, the density before perturbation is applied, we have
\begin{equation}
d\,\Delta V=-\Delta V\,\frac{dn}{n_0}.
\label{eq:deltan2}
\end{equation}

The work to change $\Delta V$ by $d\,\Delta V$ is just
\begin{equation}
d\,\Delta W=-p\,d\,\Delta V=(p_0+\delta p)\,\Delta V\,\frac{dn}{n_0}.
\label{eq:deltan3}
\end{equation}
Then the net work to change the local density from $n_0$ to $n$ is
\begin{equation}
\Delta W=\Delta V\,\int_{n_0}^n\frac{dn}{n_0}(p_0+\delta p).
\label{eq:deltan4}
\end{equation}
Using as the integration variable the local change in density $\delta n\equiv n-n_0$, $ dn = d\delta n$. Assuming an adiabatic change in pressure, $p-p_0\equiv\delta p = mc_S^2\,\delta n$, with $c_S$ the adiabatic sound speed and $p_0$ the uniform background pressure, we have
\begin{equation}
\Delta W=\Delta V\,\int_0^{\delta n}\frac{d\delta n'}{n_0}(p_0+mc_S^2\,\delta n')=\Delta V\,\left(\delta n\frac{p_0}{n_0}+ mc_S^2\,\frac{(\delta n)^2}{2\,n_0}\right).
\label{eq:deltan5}
\end{equation}
Replacing the local volume $\Delta V$ by $d^3{\mathbf r}$, we have for the total stored energy
\begin{equation}
W=\int d^3{\mathbf r}\,\left(\delta n\frac{p_0}{n_0}+ mc_S^2\,\frac{(\delta n)^2}{2\,n_0}\right)\simeq\frac{mc_S^2}{2}\int d^3{\mathbf r}\,n_0 \left[\frac{\delta n({\mathbf r})}{n_0}\right]^2.
\label{eq:deltan6}
\end{equation}
In eq.~\ref{eq:deltan6}, since the background pressure $p_0$ and density $n_0$ are spatially uniform, eq.~\ref{eq:deltan1} requires that the term linear in $\delta n$ vanish. For simplicity, we ignore the spatial variation of the sound speed $c_S$ and background density $n_0$ in the region of the perturbation. Defining the energy per particle $W_1$ in terms of the mean square fractional density perturbation, we have finally
\begin{equation}
W_1=\frac{mc_S^2}{2}\left\langle\left[\frac{\delta n({\mathbf r})}{n_0}\right]^2\right\rangle.
\label{eq:deltan7}
\end{equation}
The same result can be obtained by finding the rate of change of the total kinetic energy $K$ from the dissipationless equation of motion $n_0 m\partial_t{\mathbf v}=-\nabla\delta p$, which yields $\partial_t(K+W)=0$, so that $W$ is the effective potential energy. Note that for a unitary Fermi gas, $mc_S^2=10\,E_1/9$ from eq.~\ref{eq:1.4TP}, below. As discussed in the main text, for a sinusoidal perturbation with a 20\% amplitude, $W_1\simeq 0.01\,E_1$ is negligible.

\subsubsection{Unitary Fermi gas thermodynamics}
For the unitary Fermi gas, universality~\cite{HoUniversalThermo} requires that the pressure $p$ and the energy density ${\cal E}$ are functions only of the density and temperature, related by  $p=2\,{\cal E}/3$. Dimensional analysis then shows that the energy density takes the simple form
\begin{equation}
{\cal E}=\frac{3}{5}n\,\epsilon_F(n)\,f_E(\theta)\equiv n\,E_1,
\label{eq:1.1TP}
\end{equation}
where $E_1$ is the energy per particle and $\theta\equiv T/T_F$ is the reduced temperature with $T_F$ the local Fermi temperature. For a balanced 50-50 mixture of two spin components of total density $n$, the local Fermi energy is  $k_BT_F=\epsilon_F(n)=mv_F^2/2=\hbar^2(3\pi^2n)^{2/3}/(2m)$. The universal function $f_E(\theta)$ has been measured by Ku et al.,~\cite{KuThermo}, which determines all of the thermodynamic properties. The pressure is then
\begin{equation}
p=\frac{2}{5}n\,\epsilon_F(n)\,f_E(\theta).
\label{eq:1.2TP}
\end{equation}
The entropy density takes a similar form
\begin{equation}
s=nk_B\,f_S(\theta)=n s_1(\theta)\equiv n s_1,
\label{eq:1.3TP}
\end{equation}
where $s_1$ is the entropy per particle and $f_S(\theta)$ can be determined from  $f_E(\theta)$.

The adiabatic sound speed eq.~\ref{eq:cS} is easily obtained from eq.~\ref{eq:1.2TP}, as eq.~\ref{eq:1.3TP} requires constant $\theta$ for constant $s_1$,
\begin{equation}
m c_S^2=\left(\frac{\partial p}{\partial n}\right)_{\!\!\theta}=\frac{2}{3}\,\epsilon_F(n)\,f_E(\theta)=\frac{10}{9}\,E_1,
\label{eq:1.4TP}
\end{equation}
where the last form on the right follows from eq.~\ref{eq:1.1TP}.
With $\epsilon_F(n)=m v_F^2/2$, eq.~\ref{eq:1.4TP}  yields
\begin{equation}
c_S^2=\frac{v_F^2}{3}\,f_E(\theta).
\label{eq:adiabsound}
\end{equation}
\begin{figure}[htb]
\begin{center}\
\includegraphics[width=3.5in]{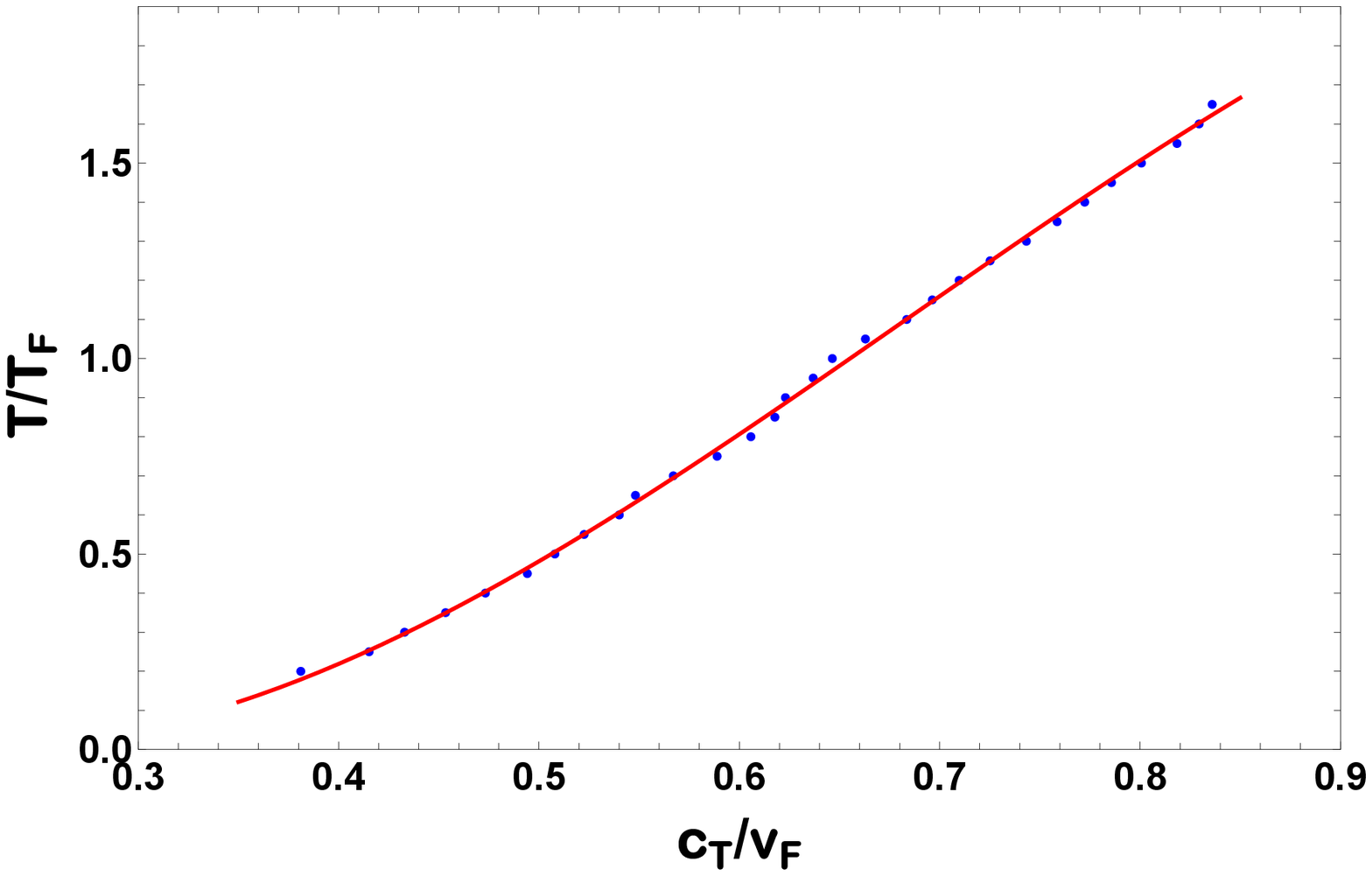}
\end{center}
\caption{Reduced temperature $\theta =T/T_F$ versus isothermal sound speed $\tilde{c}_T\equiv c_T/v_F$. For $\theta>0.25$, $\theta$ monotonically increases with $\tilde{c}_T$, showing that the fitted $\tilde{c}_T$ can be used as a thermometer to determine $\theta$ in the normal fluid region. The blue dots are obtained from the equation of state of ref.~\cite{KuThermo} The red solid curve shows the fit with a cubic polynomial, $\theta(\tilde{c}_T)=0.478 - 4.410\, \tilde{c}_T + 11.688\, \tilde{c}_T^2 - 5.711\, \tilde{c}_T^3$.
\label{fig:soundspeed}}
\end{figure}

The isothermal sound speed, eq.~\ref{eq:cT}, is easily determined from eq.~\ref{eq:1.2TP}, with $p=p[n,\theta(n,T)]$,
\begin{equation}
c_T^2=\frac{1}{m}\left(\frac{\partial p}{\partial n}\right)_{\!\!T}=\frac{v_F^2}{3}\,\left[f_E(\theta)-\frac{2}{5}\,\theta f_E'(\theta)\right].
\label{eq:isothermsound}
\end{equation}
where the $\partial_\theta f_E(\theta)\equiv f_E'(\theta)$.

The heat capacity per particle at constant volume takes a simple form. Using eq.~\ref{eq:1.1TP},
\begin{equation}
c_{V_1}=T\left(\frac{\partial s_1}{\partial T}\right)_{\!\!n}=\left(\frac{\partial E_1}{\partial T}\right)_{\!\!n}=\frac{3}{5}k_B f_E'(\theta).
\label{eq:2.1TP}
\end{equation}

Eq.~\ref{eq:1.3S} then determines the ratio $c_{P_1}/c_{V_1}=c_S^2/c_T^2$ from eqs.~\ref{eq:adiabsound}~and~\ref{eq:isothermsound},
\begin{equation}
\frac{c_{P_1}}{c_{V_1}}=\frac{f_E(\theta)}{f_E(\theta)-\frac{2}{5}\theta\,f_E'(\theta)}.
\label{eq:beta4}
\end{equation}

Finally, Eqs.~\ref{eq:2.1TP}~and~\ref{eq:beta4} determine
\begin{equation}
\frac{1}{c_{V_1}}-\frac{1}{c_{P_1}}=\frac{1}{k_B}\frac{2}{3}\frac{\theta}{f_E(\theta)},
\label{eq:DeltaInvC}
\end{equation}
which appears in the sound diffusivity Eq.~\ref{eq:D1}. The right hand side is just $4/15\,nT/p$, as obtained previously~\cite{MZSound}.

\subsection{Data Analysis Details}

\subsubsection{Determination of the box force}
\label{subsec:boxforce}
We find the force arising from the confining potential along one axis $z$, using the measured density profiles $n_0(z)$. We ignore the variation of the density along the line of site and find $n_0(z)$ from the spatially integrated column density, which is obtained from absorption images. The box potential is easily found in the local density approximation from the local chemical potential, $\mu(z)$, where $\mu(z)+U_0(z)=\mu_G$, with $\mu_G$ the global chemical potential. Then,
\begin{equation}
U_0(z)=\mu_G-\mu(z)=\mu_G-\epsilon_F[n_0(z)]\,f_{\mu}[\theta(z)].
\label{eq:mu1}
\end{equation}
Here, $f_\mu(\theta)$ is a dimensionless universal function of the reduced temperature $\theta$, which determines $\mu(n,\theta)$ in terms of the local Fermi energy $\epsilon_F(n)$. $f_\mu(\theta)$ has been precisely measured~\cite{KuThermo}. The reduced temperature $\theta(z)=T_0/T_F(n)=\theta_0/[\tilde{n}_0(z)]^{2/3}$, where we determine $\theta_0=T_0/T_F(n_0)$ from the fitted isothermal sound speed $c_T$ and $\tilde{n}_0(z)=n_0(z)/n_0$, with $n_0$ the central density, which occurs at $z\equiv z_{\rm max}$.  The global chemical potential is then $\mu_G=\epsilon_F(n_0)\,f_{\mu}(\theta_0)$, so that $U_0(z_{\rm max})=0$ by construction. It is convenient to find $\tilde{U}_0(z)=U_0(z)/\epsilon_F(n_0)$, with $\tilde{\mu}_G=f_{\mu}(\theta_0)$. Then,
\begin{equation}
\tilde{U}_0(z)=f_{\mu}(\theta_0)-[\tilde{n}_0(z)]^{\,2/3}\,f_{\mu}\!\left(\theta_0/[\tilde{n}_0(z)]^{\,2/3}\right).
\label{eq:mu2}
\end{equation}

\begin{figure}[htb]
\includegraphics[width=5.5in]{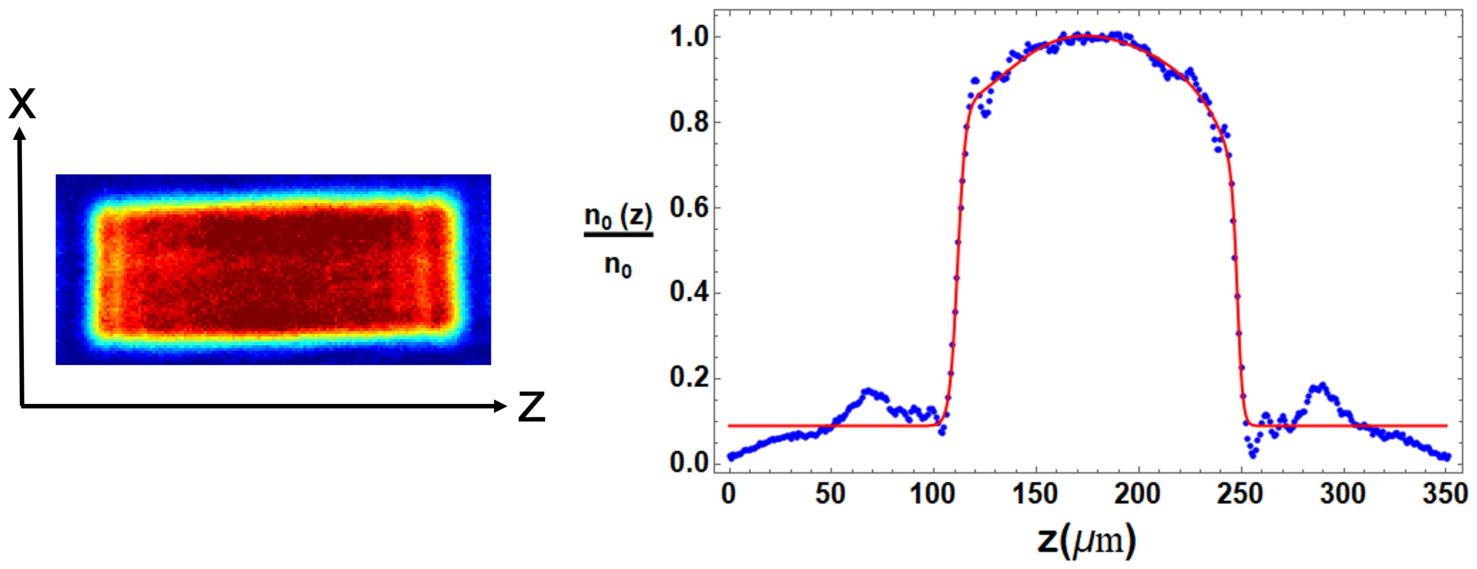}
\caption{Column Density $\tilde{n}(x,z)$ and 1D background density $n_0(z)$ (blue dots). Red curve: Fit of eq.~\ref{eq:fit}.
\label{fig:backgroundfit}}
\end{figure}
To evaluate eq.~\ref{eq:mu2}, we first fit $\tilde{n}_0(z)=n_0(z)/n_0$ with an analytic function,
\begin{equation}
h(z)= \frac{\tanh[(z - z_{10})/w_1] -
    \tanh[(z - z_{20})/w_2]}{2}\,\sum_n a_n\,z^n.
\label{eq:fit}
\end{equation}
The difference of the $\tanh$ functions produces a top-hat shape of nominal width $z_{20}-z_{10}$ and slopes on each side determined by $w_1$ and $w_2$. The flat top is modulated by the multiplying polynomial.  Fig.~\ref{fig:backgroundfit} shows a typical fit using a fifth order polynomial. The density offset arises from atoms trapped outside the box, in between the repulsive sheets and the magnetic confining potential arising from the bias magnetic field. For finding the box potential from eq.~\ref{eq:mu2}, this offset is subtracted so that the density smoothly vanishes at the walls of the box and the peak density is scaled to 1. The central 3D density $n_0$ is determined as described below in \S~\ref{sec:CentralDensity}.

\begin{figure}[htb]
\begin{center}\
\includegraphics[width=3.5in]{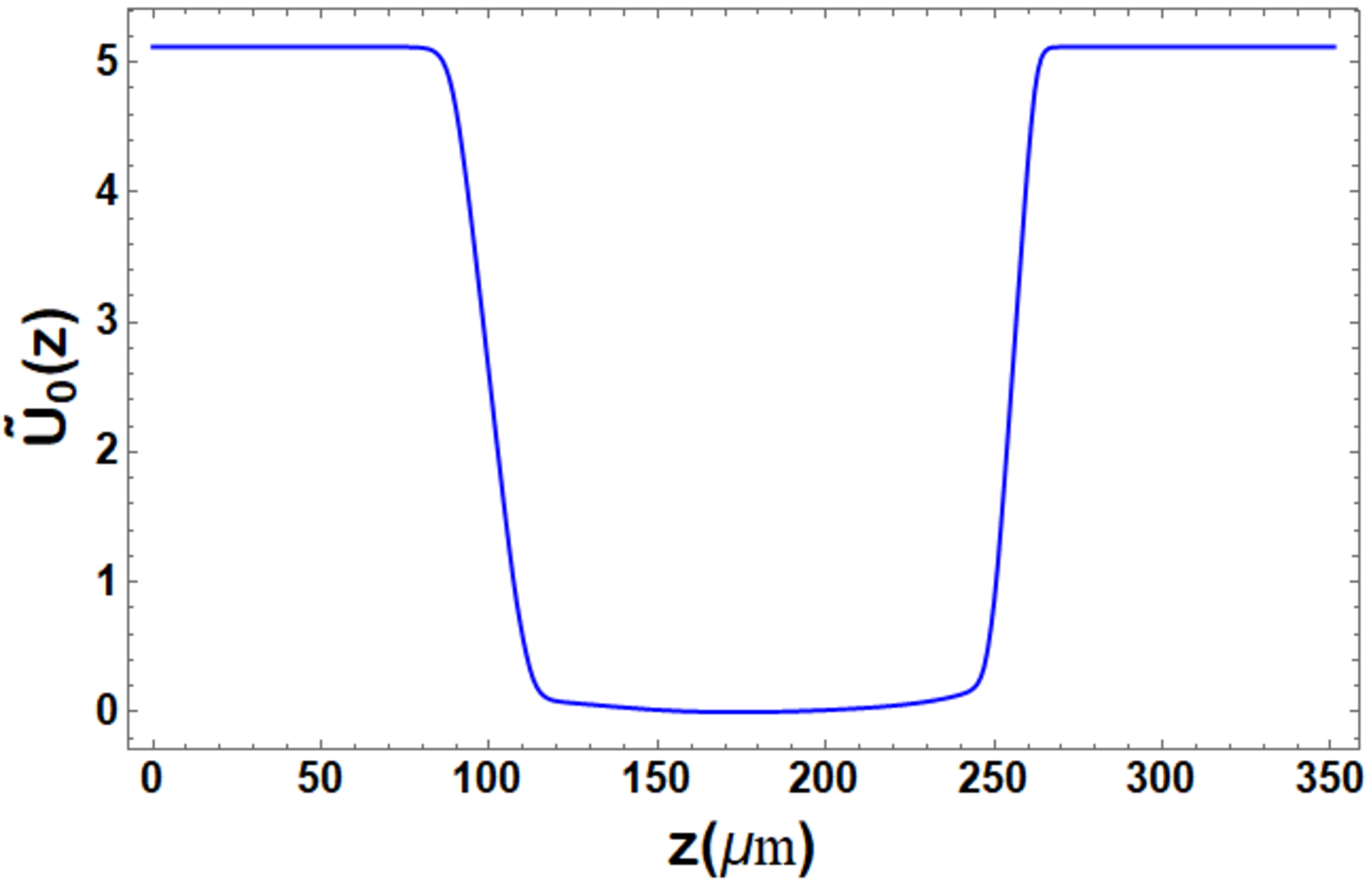}
\end{center}
\caption{Box potential in units of local Fermi energy $\epsilon_F(n_0)$ for the central density $n_0$. The potential is determined from the measured background density $n_0(z)$ using eq.~\ref{eq:mu2}. Note that the curvature at the bottom of the box potential energy arises from curvature in the bias magnetic field, which produces a small confining harmonic potential.  \label{fig:box}}
\end{figure}

The reduced temperature $\theta_0=T_0/T_F(n_0)$ is determined from the isothermal sound speed $c_T=\omega_T/q$, where the frequency $\omega_T$ is one of the fit parameters and $q$ is the measured wavevector for the $\delta n(q,t)$ data, see \S~\ref{sec:Analytic}. Using $\theta_0$, eq.~\ref{eq:mu2} yields the box potential profile, Fig.~\ref{fig:box}. The box potential then determines the corresponding force $-\epsilon_F(n_0)\,\partial_z\tilde{U}_0(z)$ for use in eq.~\ref{eq:2.5S}.
In our experiments, where $\epsilon_F(n_0)\simeq 0.2\,\mu$K, the box depth $U_0\simeq 1.0\,\mu$K.

\subsubsection{Determination of the central density}
\label{sec:CentralDensity}

The central 3D-density $n_0$ is used to find the central Fermi energy, corresponding Fermi temperature $T_F$, and Fermi speed $v_F$, which determines the reduced temperature $T/T_F$ from the measured sound speed $c_T$ using the known equation of state~\cite{KuThermo}.  The reduced temperature then determines the thermodynamic properties of the sample.
\begin{figure}[htb]
\centering\
\includegraphics[width=2.5in]{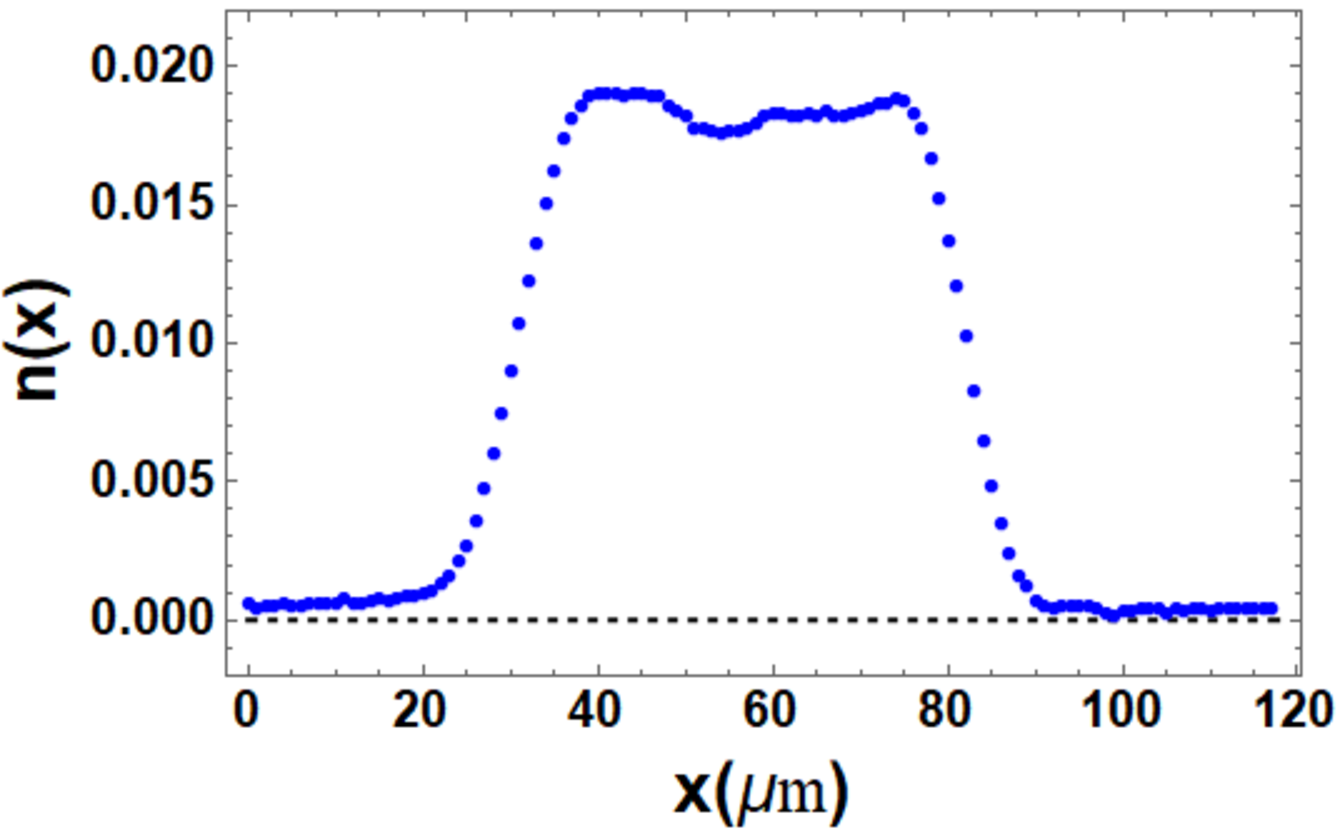}\hspace{0.25in}\includegraphics[width=2.5in]{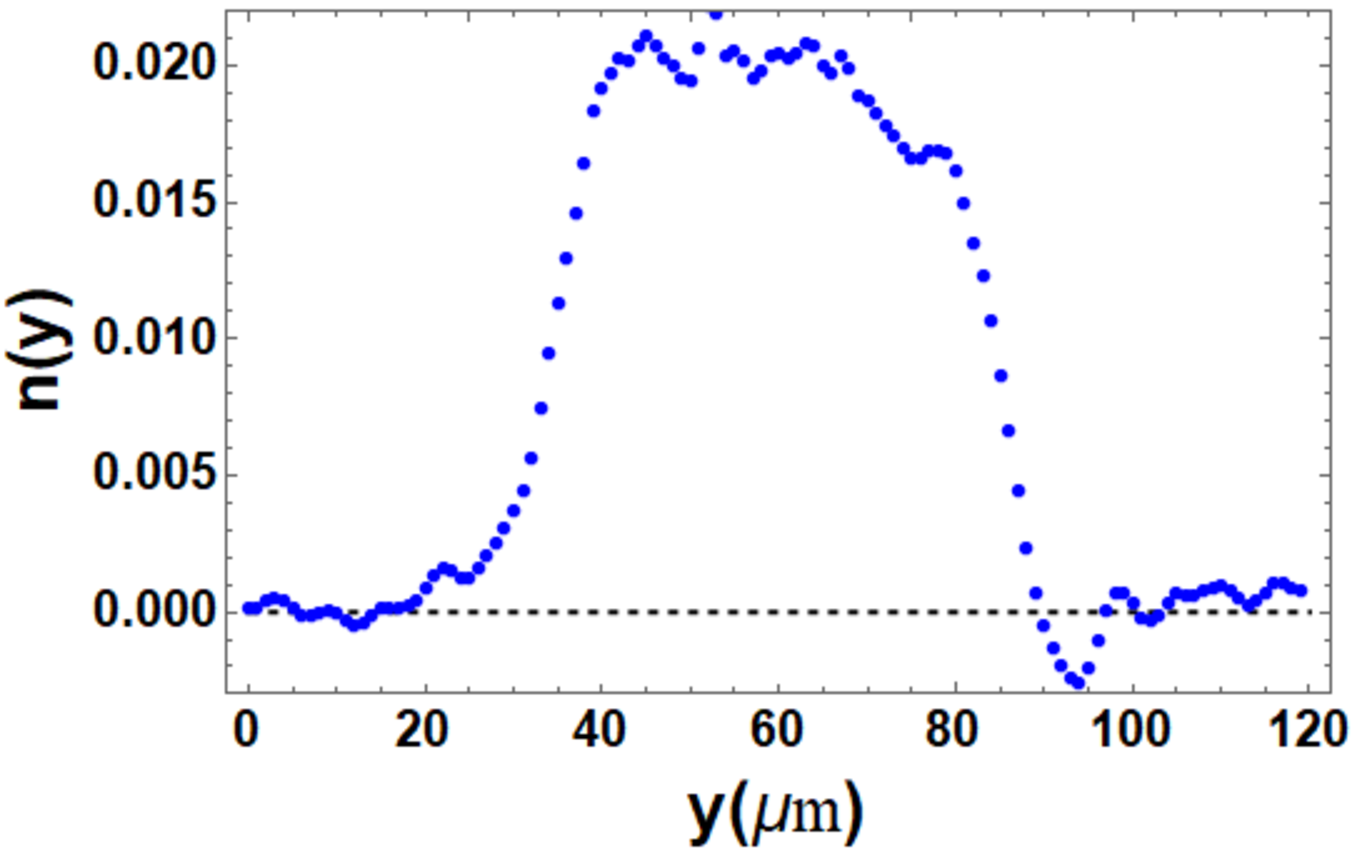}
\caption{Density profiles along the $x$ and $y$ directions of the box potential.  \label{fig:densityxy}}
\centering\
\end{figure}

To find $n_0$, we observe the trapped cloud along the $x$ and $y$ axes with two cameras, Fig.~\ref{fig:densityxy}. In this way, we measure the two-dimensional column densities $\tilde{n}(z,x)=\int_{-\infty}^\infty\! dy\,n(x,y,z)$ and $\tilde{n}(z,y)=\int_{-\infty}^\infty\! dx\, n(x,y,z)$, for each spin state, where $z$ denotes the long axis of the box potential and $x$ denotes the direction of the bias magnetic field. For our experiments, the typical box dimensions are $\Delta x=52\,\mu$m, $\Delta y=50\,\mu$m
and $\Delta z=150\,\mu$m. The curvature of the bias magnetic field produces a harmonic confining potential $\propto z^2$, which causes a noticeable variation of the density over $150\,\mu$m. In contrast, the confining potential $\propto y^2$ produces a much smaller variation of the density over $50\,\mu$m along the $y$ axis. Further, the number of atoms trapped outside the box along $x$ and $y$ is negligible. We note that the measured $n(y)$ is distorted on the right side. This is an artifact of the imaging path for the vertical camera, which is collinear with the vertically projected beams that form the sides of the box potential. We assume that the true shapes are nearly identical.

The one-dimensional density that we analyze in the experiments as a function of time, is obtained by integrating the measured 2D-column density over a limited central region along $x$, where the density is slowly varying,
\begin{equation}
n(z)=\int_{x_1}^{x_2} dx\,\tilde n(z,x).
\label{eq:1Ddensity}
\end{equation}
To estimate the 3D density, we assume that $n(x,y,z)$ approximately factors, as it would in a true 3D box potential,
\begin{equation}
n(x,y,z)\simeq \tilde{n}(x,z)\,n(y).
\label{eq:3Ddensity}
\end{equation}
We normalize $\int_{-\infty}^\infty\!dy\,n(y) =1$, so that $\int_{-\infty}^\infty\!dy\,n(x,y,z)=\tilde{n}(x,z)$ as it should.  The normalized 1D density $n(y)$, averaged near the center of the box $y=y_c$, is essentially the inverse of the box length $L_y$ along $y$, as it would for a true box potential.   We measure
\begin{equation}
n(y)=\frac{\int_{z_1}^{z_2} dz\,\tilde{n}(z,y)}{\int_{-\infty}^\infty\!dy\int_{z_1}^{z_2} dz\,\tilde{n}(z,y)},
\label{eq:norm}
\end{equation}
where $\tilde{n}(z,y)$ is the column density measured by the camera oriented along the $x$-axis. $n(y)$ satisfies $\int_{-\infty}^\infty\!dy\,n(y) =1$ for any choice of $z_1$ and $z_2$. We take $z_1$ and $z_2$ in the central region of the cloud, where the density is nearly uniform, as used to measure the Fourier transform $\delta n(q,t)$.

Averaging the column density near the center, $x_c,z_c$, where the 2D density is nearly uniform, we obtain the total central density $n_0=2\,\tilde{n}(x_c,z_c)\,n(y_c)$. For a single spin state, typical values are $n(y_c)=0.0204/\mu{\rm m}=204/$cm, i.e., $L_y=49.0\,\mu$m, and $\tilde{n}(x_c,z_c)=1.10\times10^9/{\rm cm}^2$, which yields $n_0/2=2.24\times 10^{11}/{\rm cm}^3$. From $n_0$ we find the Fermi speed $v_F$. This in turn determines the reduced temperature $\theta(c_T/v_F)$, Fig.~\ref{fig:soundspeed}, where $c_T=\omega_T/q$ is determined from the fit to $\delta n(q,t)$.

\subsubsection{Effect of Density Variation on the Measured Transport Coefficients}
\label{sec:density}
Figs.~3-5 of the main text show the measured transport coefficients, where the error bars denote the statistical errors from the $\chi^2$ fits, which we find from the error matrix. The transport coefficients $\alpha_\eta$ and $\alpha_\kappa$, and $2\,a/q^2$, are determined by eqs.~\ref{eq:frequencies}, which do not explicitly depend on the density.

However, the measured decay rates are inherently averages over the sample. To estimate the effect of the density variation, we consider the high temperature limit, where  $c_T\propto\sqrt{T}$ is independent of density, as are the transport properties, $\eta$ and $\kappa$, which are $\propto T^{3/2}$. The decay rates then scale inversely with density, $\gamma_\eta\equiv \gamma_\eta(0)\,n_0/n$, and $\gamma_\kappa\equiv \gamma_\kappa(0)\,n_0/n$, so that the decay rates are larger in the low density regions compared to the center, where $n(0)=n_0$.

In a simple model, we can average the exponential decay factors with a normalized density profile for the region measured in our $150\,\mu$m boxes, where the density variation over the central 100 microns is $\simeq 10$\%. We use  the central values $\gamma_\eta(0)$ and $\gamma_\kappa(0)$ as fit parameters, since these correspond to the density $n_0$ that determines $T/T_F$ in the figures. These fit parameters  are adjusted so that the average decay factors agree with the measurements. We find that $\gamma_\eta(0)$ and $\gamma_\kappa(0)$ are shifted downward by 5\% compared to the measured values.  These results are confirmed by numerical modeling of $\delta n(z,t)$ with Eqs.~\ref{eq:2.5S}~and~\ref{eq:1.9Sc}, where we find a downward shift of 3\% for $\gamma_\eta(0)$ and 6\% for $\gamma_\kappa(0)$. We also compute the corresponding average for the density profile $n(y)$ along the line-of-sight direction, where we cannot choose the central region. We divide the density $n(y)$ into 30 segments, find $\delta n(q,t)$ for each segment, and sum the density weighted decay curves,  yielding comparable shifts.

From these estimates, we see that the corrected transport parameters, corresponding to the central density $n_0$ and the given $T/T_F$, are systematically shifted downward, compared to the given measured values, by at most $5$\%.


\begin{thebibliography}{30}
\bibitem [{Bul()}]{Bulk}%
  \BibitemOpen
  \href@noop {} {}\bibinfo {note} {For a unitary Fermi gas, the bulk viscosity
  $\xi_B$ has been measured~\cite{ElliottScaleInv} and found to be negligible
  compared to the shear viscosity, consistent with predictions that $\xi_B=0$
  for scale invariant
  systems~\cite{SonBulkViscosity,StringariBulk}.}\BibitemShut {Stop}%
\bibitem [{The()}]{ThetaError}%
  \BibitemOpen
  \href@noop {} {}\bibinfo {note} {We give $\theta=T/T_F$ for the average
  central density $n_0$~\cite{SupportOnline}}\BibitemShut {NoStop}%
\end{thebibliography}
\end{document}